\documentclass{article}
\usepackage{latexsym,amsmath, amssymb,ulem}

\newenvironment{acknowledgements}{
\noindent {\it
    Acknowledgements:}}{\\[1ex]}
\newenvironment{mytitle}{\noindent }{\\}
\newenvironment{myauthor}{\\ \noindent}{\\}
\newenvironment{keywords}{Keywords: }{\\}
\newcommand{\new}{\newcommand}
\newcounter{letter}

\newcommand{\bracket}[1]{\left\langle #1 \right \rangle}
\newcommand{\parens}[1]{\!\left( #1 \right)}
\newcommand{\sbrace}[1]{\!\left[ #1 \right]}

\newcommand{\abs}[1]{\left| #1 \right|}
\newcommand{\norm}[1]{\left|\!\left| #1 \right|\!\right|}
\makeatletter
\def\subsection{\@startsection {subsection}{2}{\z@}{.8ex
plus 1ex} {1ex}{\sc}}
\makeatother
\numberwithin{equation}{section}
\newtheorem{remark}{Remark}[section]

\newtheorem{definition}{Definition}

\newcommand{\NN}{\mathbb{N}}
\newcommand{\RR}{\mathbb{R}}
\newcommand{\ZZ}{\mathbb{Z}}

\newcommand{\OO}{\mathcal{O}}
\newcommand{\tensor}{\otimes}
\newcommand{\End}{\operatorname{End}}

\new{\Ricci}{\mathrm{Ricci}}
\new{\scalar}{{\mathfrak{r}}}
\new{\Pfaff}{\mathrm{Pfaff}}
\new{\Nabla}{\bigtriangledown}
\renewcommand{\vector}{\mathbf}
\renewcommand{\vec}{\vector}
\new{\Hom}{\mathrm{Hom}}
\new{\kernel}{\mathrm{Kernel}}
\new{\grad}{\mathrm{grad}}
\new{\diverge}{\mathrm{div}}
\new{\tr}{\mathrm{tr}}
\new{\indx}{\mathrm{ind}}
\new{\str}{\mathrm{str}}
\new{\Str}{\mathrm{Str}}
\new{\kk}{{\mathfrak{k}}}
\new{\lop}{{\mathfrak{l}}}
\new{\defequals}{\stackrel{\mathrm{def}}{=}}
\new{\im}{\mathrm{Im}}
\new{\boxtensor}{\boxtimes}
\new{\barpsi}{\overline{\psi}}
\new{\barrho}{\overline{\rho}}
\new{\iso}{\cong}
\new{\vol}{\operatorname{Vol}}
\new{\gap}{{\;}}
\new{\depends}[1]{{\scriptscriptstyle{#1\!\!}}}
\new{\triplegap}{{\;\;\;\;\;}}
\new{\doublegap}{{\;\;\;}}
\new{\FF}{{\mathcal{F}}}
\new{\Alg}{{\mathcal{A}}}
\new{\bundle}{{\mathcal{E}}}
\new{\QQ}{{\mathcal{Q}}}
\new{\MM}{{\mathcal{M}}}
\new{\MMm}{{\mathcal{M}}_m}
\newcommand{\Clifford}{{\mathcal{C}}}
\new{\PP}{{\mathcal{P}}}
\new{\spinor}{{\mathcal{S}}}
\new{\tspinor}{{\mathcal{T}}}
\new{\dirac}{\textsf{\textbf{D}}}
\new{\tnm}[1]{{(#1)}}
\new{\EE}{{\mathcal{E}}}
\new{\DD}{{\mathcal{D}}}
\new{\VV}{{\mathcal{V}}}
\new{\XX}{{\mathcal{X}}}
\new{\vecspace}{{\mathbb{V}}}
\new{\Lie}{{\mathcal{L}}}
\new{\dd}{{\mathfrak{d}}}
\new{\dmeas}{{\mathbf{d}}}
\new{\Rlimit}{{\mathsf{R}}}
\new{\Flimit}{{\mathsf{F}}}
\new{\kheat}{K_{\mbox{{\scriptsize heat}}}}
\new{\twisted}{\mbox{{\scriptsize twisted}}}
\new{\be}{\begin{equation}  
}
\new{\ee}[1]{
\label{eq:#1}
\end{equation} \noindent%
}
\new{\eqa}[2]{\begin{align}
#2 \label{eq:#1}
\end{align}
\ifnum \thedraft=1 \marginpar{\scriptsize{\em{eq:#1}}}  \fi
}
\new{\pt}{\mathfrak{P}}
\new{\ptxystar}{\pt^{x*}_{y}}
\new{\ptyxstar}{\pt^{y*}_{x}}
\new{\bint}{\oint}

\renewcommand{\l}{\left}
\renewcommand{\r}{\right}

\new{\gammadot}{\dot{\gamma}}
\new{\sigmadot}{\dot{\sigma}}
\newcounter{draft} 
\ifnum \thedraft=0
\fi
\makeatletter
\def\subsection{\@startsection {subsection}{2}{\z@}{.8ex
plus 1ex} {1ex}{\sc}}
\makeatother
\new{\phone}[1]{Phone: #1}
\new{\fax}[1]{Fax: #1}
\begin{document}
%
{\bf \large \mytitle{Rigorous path integrals for supersymmetric quantum mechanics:\\ completing
  the path integral proof of the index theorem}}\\[3ex]
\myauthor{\hspace*{2in} Dana S. Fine\footnote{University of Massachusetts Dartmouth, N. Dartmouth, MA
  02747} and Stephen F. Sawin\footnote{Fairfield University,
      Fairfield, CT 06824}}
\begin{abstract}
Many introductory courses in quantum mechanics include Feynman's
time-slicing definition of the path integral, with a complete
derivation of the propagator in the simplest of cases. However,
attempts to generalize this, for instance to non-quadratic potentials,
encounter formidable analytic issues in showing the successive
approximations in fact converge to a definite expression for the path
integral. The present work describes how to carry out the analysis for
a class of Lagrangians broad enough to include the evolution, in
imaginary time, of spinors constrained to live on a Riemannian
manifold. For these Lagrangians, the successive time-slicing
approximations converge. The limit provides a definition of the path
integral which agrees with the imaginary-time Feynman propagator. With
this as the definition, the steepest-descent approximation to the path
integral for twisted $N=1/2$ supersymmetric quantum mechanics is
provably correct. These results complete a new proof of the
Atiyah-Singer index theorem for the twisted Dirac operator.
\end{abstract}
\section*{Introduction}
%
Elaborating on an argument due to Witten\cite{Witten82a},
Alvarez-Gaum\'{e}\cite{Alvarez83} evaluates path integrals for 
supersymmetric quantum mechanics (SUSYQM) using what has become a
familiar argument, the short version of which  is that for
cohomological quantum theories stationary phase is exact. Friedan and
Windey\cite{FW84} extend this to a ``twisted'' version.

In slightly more detail, the path integral in question is 
\[
 \int  e^{-S(\sigma, \Psi,
      \Psi^\dagger, t)} \, \DD \Psi^\dagger \DD \Psi \DD \sigma \,  d\psi^\dagger d\psi dx
  \]
  where the action $S$ is the time integral of the SUSYQM Lagrangian, and the integral is over the space of paths which start at a pair $(y, \psi_y)$,
  consisting of a point $y$ in a Riemannian manifold $M$ and an associated
  spinor $\psi_y$, and 
  after a time interval of length $t$ end at the pair $(x,
  \psi_x)$. The paths  
consist of a standard path $\sigma$ in $M$ and, at
each point of $\sigma$, a spinor $\Psi$  and a dual spinor
$\Psi^\dagger$. 
A close
reading of Alvarez-Gaum\'{e}'s 
treatment of the Euclidean theory for SUSYQM on $M$ reveals the argument rests on two
key properties of the path integral:
\begin{enumerate}
\item
The path integral represents the heat kernel
of a Laplacian operator on a known bundle. 
\item
Steepest-descent ({\it i.e.},  the standard path integral technique of
stationary phase but in the Euclidean realm) 
  provides an approximation to the path integral valid for small
  $t$.
\end{enumerate}
Taking an appropriate supertrace of the heat kernel, 
which corresponds to taking the integral over loops rather than paths
with fixed endpoints, calculates a known topological invariant. For
instance, in the $N=1$ theory this is the Euler characteristic, while
in the $N=1/2$ theory it is the index of the Dirac
operator. Necessarily, such invariants do not depend on the time
parameter $t$ appearing in the path integral.
Thus, assuming the above 
properties hold, taking $t$ to $0$ in the
 steepest-descent approximation to the path integral over loops gives the {\em exact} value of the path
 integral, and with it the supertrace of the heat kernel. Calculating
 the small-$t$ limit of the standard expression for the steepest
 descent approximation gives an ordinary integral over $M$. Equating
 this integral with the supertrace gives a path integral proof of the
corresponding index theorem.
 In the above cases these are the
 Gauss-Bonnet-Chern theorem and the Atiyah-Singer index theorem,
 respectively.

 From a mathematical viewpoint, these elegant arguments must be taken
 to be merely heuristic, because the relevant path integrals
 themselves have not been carefully defined. Most crucially, the
 stated properties have been proven for rigorous definitions of path
 integrals only in settings much simpler than those of SUSYQM on manifolds.

  The following, which may be thought of as an explanation of some
  analytic details missing from the derivations
  in~\cite{Witten82a},~\cite{Alvarez83} and~\cite{FW84}, describes the
  rigorous mathematical construction, based on Feynman's original
  time-slicing definition~\cite{Feynman42}, of path integrals for a large class of
  Lagrangians; namely, those corresponding to generalized Laplacians
  on bundles on manifolds. The class is large enough to include the
  Lagrangian for the twisted version of $N=1/2$ SUSYQM of Friedan and
  Windey\cite{FW84} from which they derive the Atiyah-Singer index
  theorem for the twisted Dirac operator, which includes the others a
  special cases. The present construction of the path integral agrees
  with the heat kernel of the generalized Laplacian, so Property 1
  holds true. In the special case of twisted $N=1/2$ SUSYQM on loops,
  bounds on the error between the time-slicing approximation based on
  a given partition of $t$ and the path integral suffice to
  interchange the fine partition and the small-$t$ limit. This
  interchange of limits leads to a proof of the validity of the
  steepest descent approximation, which is Property 2, to sufficiently
  high order in $t$ to complete a new proof 
  of the Atiyah-Singer index theorem for the twisted Dirac operator.

Starting from the Lagrangian, several choices go into defining the
time-slicing approximation to the path integral. These include choices
of Riemann sum approximations to the action, and terms which might
explicitly depend on $\hbar$. While the various choices should be
equivalent in the sense of the existence of a fine-partition limit,
and its value, the argument in Sect.~\ref{ss:convergence} that the
approximate path integrals converge to the heat kernel depends on a
particular estimate which constrains the choice of time-slicing
approximation.  Remark~\ref{rm:correction} discusses this is some
detail, in particular identifying many choices leading to the same
limiting path integral. Presumably, choices satisfying the additional
constraint make the rate of convergence manifestly faster.
\subsection*{Other approaches to rigorous path integrals and index theorems}
The elegant heuristic arguments have inspired a variety of
mathematically rigorous approaches.
Bismut\cite{Bismut84a,Bismut84b} uses stochastic techniques with the
heat equation to give a proof of the index theorem in the spirit of
the physics argument. Getzler\cite{Getzler86a,Getzler86b}, who does
not directly construct path integrals, gives an index theorem proof
using the theory of pseudo-differential operators to provide the
estimates suggested by these arguments.  Rogers\cite{Rogers87} uses
stochastic techniques to construct an explicit supersymmetric path
integral for the heat kernel on manifolds whose Riemannian metric is
Euclidean outside of a bounded region. This suffices to reproduce the
path integral proof of the GBC theorem for arbitrary compact
manifolds, since the argument only depends on the short-time behavior
of the restriction of the heat kernel to the diagonal.  In later
work\cite{Rogers92a,Rogers92b}, she extends these techniques to prove
the twisted Hirzebruch index theorem, from which follows the full
index theorem.  Andersson and Driver\cite{AD99} use stochastic
techniques to construct a version of the bosonic path integral on
curved space.

The present paper describes an argument, which Fine and
Sawin\cite{FS17} presents in more technical terms, whose innovation
is to make rigorous 
Feynman's  time-slicing procedure in constructing the
supersymmetric path integral, thereby representing the heat kernel for the
any generalized Laplacian on an arbitrary compact Riemannian manifold
as a path integral, and to obtain, from this representation, its
short-time approximation on the diagonal in the special case of
$N=1/2$ SUSYQM.

\section{The action for a generalized Laplacian on the Grassman algebra of
  a vector bundle and a corresponding time-slicing approximate kernel}
\subsection{Generalized Laplacians as Hamiltonians and the
corresponding Lagrangians} \label{sec:tame}
Let $M$ be a Riemannian manifold, with metric $g$, and let $\VV$ be a
vector bundle over $M$. Let $\nabla$ be a generalized Laplacian; that
is, a second-order  a second-order elliptic
   operator on sections of $\VV$ 
   which in local coordinates has the form
   \[
   \Delta = g^{ij} \frac{\partial^2}{\partial x_i \partial x_j} + A^i
   \frac{\partial}{\partial x_i} + B,
   \]
   with $A^i$ and $B$ 
   valued in   $\operatorname{Matrix}_{n,n}$. The goal is to construct
   path integrals for quantum mechanics on $M$ with (imaginary-time) Hamiltonian given
   by $\Delta$. Generalized Laplacians include Hamiltonians with
   non-trivial potentials, while  generalizing to vector
   bundles allows for additional structure such as spinors.

   If the data
   $(M,g,\VV,\Delta)$ are not all smooth, or $M$ is not compact, 
 require that the data be tame in the following
 sense:\footnote{Noncompact manifolds arise in the argument for the
   convergence of the approximate path integrals, because the key
   properties prove to be local. The noncompact manifold $\RR^m$
   provides the easiest setting in which to
   formulate these properties. Noncompact manifolds, again $\RR^m$,
   appear in Sect.~\ref{sss:diagonal-neighborhood} which extends local data to ultimately prove  the
   steepest descent expression gives the correct small-$t$ behavior of
   the path integral. This is also the case Rogers
   treats using stochastic quantization~\cite{Rogers92a,Rogers92b}.}
\begin{definition}
 \label{def:tame}
   An atlas of charts for $\VV$ over $M$ is 
   \emph{tame} if  
   \begin{itemize}
   \item All derivatives of $g$ and $g^{-1}$ of order $0 \leq k \leq
     6$, expressed in the coordinates of each chart, 
       are uniformly bounded in the supremum norm on
     all charts. 
   \item There is a $D_0>0$ such that the ball of radius $D_0$ around any point is contained in a single chart.
   \end{itemize}
   The tuple $(M,g,\VV)$ is \emph{tame} if it admits a tame atlas.  
   If  $\Delta$ is a 
   generalized Laplacian,
   and if there is a tame atlas so that the 
   derivatives of order $0 \leq k \leq 2$ of  $A^j$ and $B$ 
   in all charts  are uniformly bounded in the supremum norm, then say that
   $(M,g,\VV,\Delta)$ is 
   \emph{tame.} 
\end{definition}
If $M$ is compact and all the data is smooth, then
$(M,g,\VV,\Delta)$ is necessarily tame. Tameness
is a technical restriction ensuring in more general settings that the
manifold, metric, bundle and 
generalized Laplacian are sufficiently smooth to apply the convergence
arguments of Sect.~\ref{s:convergence_broad}.  

In passing to the imaginary-time formalism, the time evolution
operator associated to the Hamiltonian $\Delta$ becomes the heat
operator $e^{-t\Delta/2}$ taking an initial configuration $f_0(x)$
to the corresponding solution $f(x,  t)$ of the heat equation
\[
\frac{1}{2} \Delta f = \frac{\partial f}{\partial t}.
\]
Indeed, the heat equation is the imaginary-time Schr\"{o}dinger equation
governing the evolution of the wave-function $f(x, t)$ from its
initial value $f_0(x)$. 
Here $x \in M$ is a point of $M$ while $f(x,t)$ is a point in the
fiber of $\VV$ over $x$. In local 
coordinates, these are specified by the coordinates $x^\mu$
of a point in $R^n$ and the components
$f^a (x, t)$ of a vector, relative
to a particular basis $e_a(x)$, in the vector space defining $\VV$.
In the cases of most interest, the bundle will actually have the form
$\XX = \Lambda \VV$ for some other bundle $\VV$ and the sections can
be written as $f(x, \psi, t)$ for $\psi$ a 
Grassman-valued, or anti-commuting, section of $\VV$; see
Sect.~\ref{ss:grassman} below. 

Berline, Getzler and 
	Vergne \cite{BGV04} observe that every generalized Laplacian can be
	written locally as
	\be
	\Delta^\VV = g^{ij} \sbrace{\nabla_{\partial_i}^\VV
          \nabla_{\partial_j}^\VV -
          \Gamma_{ij}^k\nabla_{\partial_k}^\VV} -V, 
	\ee{delta-def}
	where $\nabla^\VV$ is the covariant derivative defined by a connection on $\VV,$
	$\nabla^{\text{LC}}_{\partial_i}(\partial_j)=	\Gamma_{ij}^k \partial_k$ defines the Christoffel symbols for the
	Levi-Civita connection on the tangent bundle, and $V$ is a section of
	$\End(\VV).$ (In local coordinates, relative to the basis
        $e_a(x)$,  $V$ is a matrix-valued
	function $V^a_{\: b}(x)$.)

Let $\sigma$ denote a path in $M$ with parameter $s$, $\Psi$ a lift of
$\sigma$ to $\VV^*$, the dual vector bundle, and $\Psi^\dagger$ a lift
to $\VV$. The Lagrangian corresponding to $\Delta$ is then 
\[
L(\sigma, \Psi, \Psi^\dagger, s) = \frac{1}{2}\parens{\dot{\sigma},\dot{\sigma}} + i
\bracket{\Psi ^\dagger, \nabla^\VV_{\dot{\sigma}} \Psi} 
-\frac{i}{2}\bracket{V \Psi^\dagger,  \Psi}.
\]
In local coordinates, with $\nabla^{\VV}_{\partial_i}e_b = A_i^a
e_a$, this is
\[
L(\sigma, \Psi, \Psi^\dagger, s) = \frac{1}{2}
g_{i j} \dot{\sigma}^i \dot{\sigma}^j + i \parens{\Psi^\dagger}_a
\parens{\dot{\Psi}^a + A_{i b}^a \dot{\sigma}^{i}\Psi^b} - \frac{i}{2}
V_{\: b} \parens{\Psi^\dagger}_a \Psi^b.
\]
Here the dot refers to the derivative with respect to $s$,  $\sigma^i$
is evaluated at $s$, while $g_{ij}$, 
$\Psi^\dagger$, $\Psi$, $A_{i b}^a$ and $V^a_{\: b}$ are evaluated at
$\sigma(s)$. The action is just the time integral of the Lagrangian,
\[
S(\sigma, \Psi, \Psi^\dagger, t) = \int_0^t L(\sigma, \Psi,
\Psi^\dagger, s) \, ds.
\]
\subsection{Feynman's time-slicing approximation to the path integral
as a kernel}
Consider the case of the usual Laplacian on functions (corresponding
to $\VV = M \times \RR$ and $V=0$);
that is, the Euclidean version of the Hamiltonian for a bosonic
particle moving in $M$. Its heat kernel  is a function
$\kheat(x,y; t)$ of a pair of points of $M$ and the time parameter
$t$ determined by
\be
\parens{e^{t\Delta /2 } f_0 }(x, t) = \int \kheat(x,y; t) f_0(y) \,
dy,
\ee{Kf}
where the integral is with respect the Riemannian volume form on
$M$. In other words,  $\kheat$ implements the heat operator as an integral
kernel. 
The path integral
\[
\int e^{-\int_0^t L \, ds} \dd \sigma,
\]
where $\sigma : [0,t] \to M$ is a path from $y$ to $x$, and
$L(\sigma, \dot{\sigma}, s)$ is the Lagrangian, should be equal to
$\kheat (x,y;t)$.  (Here $\dot{\sigma} = \frac{d\sigma}{ds}$.)
The argument for this is Feynman's time-slicing interpretation of the
path integral: Partition $[0,t]$ into subintervals of length
$t_i$ for $i = 1,2 \ldots n$. Write the path integral as a product of $n$
 such
integrals, where $y_{i-1}$ and $y_i$ are the starting and ending
values, respectively, of the path in the  $i$th integral, and the
product is integrated over all the repeated points of $M$. In each of
these path integrals,  replace the integral of $L$ over the
subinterval of length 
$t_i$ with an approximation $\widehat{L}(y_i, y_{i-1}; t_i)
t_i$. Require
that $ \sum\widehat{L}(y_i, y_{i-1}; t_i) t_i$ be a Riemann
sum converging under refinement to $\int_0^t L \, ds$. The choice of
$\widehat{L}$ determines
an approximate heat kernel $K(x,y;t)= (2 \pi
t)^{-m/2}e^{-\widehat{L}(x,y;t)t}$ and an
approximate path integral; namely, the kernel product
\be
\int e^{-\int_0^t L \, ds} \dd \sigma \approx \int K(x,y_{n-1}; t_n) K(y_{n-1},
y_{n-2}; t_{n-1}) \cdots K(y_1, y; t_1) \, dy_{n-1} \cdots dy_1
\ee{time_slicing}
of $n$ copies of
$K$. If $K$ happens to have the semigroup
property, then the approximation is independent of the choice
of partition; the convergence of the approximate path integral is immediate
 in this case.  The Riemann sum requirement suggests that if $t$ itself is small enough, the
trivial partition with $n=1$ should give a good approximation to the
fine-partition limit; hence, $K$ should be close
to the actual heat kernel when $t$ is small. In the special
case $M = \RR^m$ and $V=0$,
which corresponds 
to  a free particle in
flat spacetime, defining the approximate kernel $K$ by
$\widehat{L}(x,y;t) t = \int_0^t
L(\sigma_{\mathrm{cl}},\dot{\sigma}_{\mathrm{cl}};s) \, ds$ where
$\sigma_{\mathrm{cl}}$ is the path obeying the classical equations of
motion subject to $\sigma_{\mathrm{cl}}(0)= y$ and
$\sigma_{\mathrm{cl}}(t)= x$,  happens to make $K$ exactly the heat
kernel $\kheat$. This is a semigroup, and it is immediate that the
approximations converge to a limit path integral which is
$\kheat(x,y;t)$. If $V$ is quadratic, and $M$ is still $\RR^m$, explicitly
calculating the  successive 
approximations is straight-forward, and the resulting expressions
converge to the heat kernel. However, in the general setting, the time-slicing
approximations may fail to converge as the partitions become
finer. Even if they are known to converge, there is a separate
question of whether the limiting kernel is the heat kernel. For
example, on a more general compact manifold $M$, even for the
Laplacian on functions, choosing $\widehat{L}$ analogously leads to an
approximate kernel $K$ for which, although the approximate path
integrals converge, the limiting kernel is not the heat kernel for
this Laplacian. To get the desired Laplacian requires modifying $K$ by
correction terms, which, as in physical units they enter at higher powers of
$\hbar$, may be
thought of as resolving  operator-ordering ambiguities.
\subsection{Grassman-valued variables} \label{ss:grassman}
The Lagrangian for SUSYQM refers to spinors, which are sections
of the Grassman algebra of a certain vector bundle, which can be
expressed as
functions of Grassman-valued variables as follows:
If $f(v_1, \ldots,v_n)$ is a multilinear function of $\vecspace^*$ for some
vector space $\vecspace,$ then the antisymmetrization of $f$ represents an
element of $\Lambda^n \vecspace$  To  say $\psi$ is a \emph{Grassman
  variable} valued in $\vecspace^*,$ means that the expression
$f(\psi,\ldots, \psi)$, represents that element.   More generally, write
$f(\psi)$ for a linear combination of forms of various degrees,
      {\it i.e.\/}
a {\em multiform}. If $\vecspace$ has an inner product
the \emph{Berezin Integral} $\bint f(\psi) 
 d\psi$  is the coefficient  
 of the canonical top-degree  element of $\Lambda \vecspace$ in
 $f(\psi)$.  See~\cite{MQ86} for a standard reference on Grassman
variables; \cite{Rogers92a} and~\cite{FS08} give examples relevant to
SUSYQM. 
\subsection{The action}
Suppose $(M,g, \VV,\nabla, V)$ are, respectively, a Riemannian
manifold, its metric, a vector bundle over $M$, a connection on $\VV$
and a section of $\End(\VV)$. Recall these are the ingredients required to
define a generalized Laplacian $\Delta^{\VV}$ acting on sections of $\VV$.  Let
$\XX=\Lambda \VV$, promote $\nabla$ and $V$ to a
connection and operator on $ \XX$ using
$\nabla(a \wedge b)= \nabla(a) \wedge b +  a \wedge \nabla(b)$ and
$V(a \wedge b)= V(a) \wedge b +  a \wedge V(b)$, and let $\Delta^{ \XX}$ be the generalized
Laplacian associated  to $\XX$. For each point $x \in M$ let $\psi_x$ be
a Grassman variable valued in $\VV^*_x$ so as to
 write kernels on $\XX$ as
superkernels $K(x,y,\psi_x,\psi_y).$ Here $K$ acts on
a section of $\XX,$ which is represented by a  superfunction
$f(x,\psi_x),$ as
\[(K*f)(x,\psi_x)= \int \bint K(x,y,\psi_x,\psi_y) f(y,\psi_y) d\psi_y
\, dy.\]

Let $\sigma(s)$ be a path in
$M$, let $\Psi,\Psi^\dagger$ be  Grassman variables valued in lifts of
$\sigma$ to $\VV^*$ and $\VV$   respectively, and consider the action
\be
\int \frac{1}{2}\parens{\dot{\sigma},\dot{\sigma}} + i
\bracket{\Psi ^\dagger, \nabla^\VV_s \Psi} 
-\frac{i}{2}\bracket{V \Psi^\dagger,  \Psi} ds.
\ee{action-general}

To construct a time-slicing approximate kernel, consider  a small interval of
parameter length $t$, and approximate the path connecting $x$ and $y$ by a
geodesic. This  gives $\int \frac{1}{2}\parens{\dot{\sigma},\dot{\sigma}} dt \sim
 \parens{\vec{x}_y/ t,\vec{x}_y/ t} t /2 \sim
\abs{\vec{x}_y}^2/(2 t)$, where $\vec{x}_y \in T_y M$ satisfies
$\exp_y(\vec{x}_y) = x$.
Assuming 
$\Psi^\dagger$ and $\nabla_s\Psi$ are covariantly slowly varying,
$ \int i\bracket{\Psi^\dagger, \nabla_s \Psi} ds \sim i \bracket{
  \Psi^\dagger(t_y) , \pt_y^x \Psi(t_x) -
  \Psi(t_y)}=i\bracket{\psi^\dagger_y, \pt^x_y \psi_x - \psi_y}$
and  $ \int
i\bracket{V\Psi^\dagger, \Psi} ds \sim i \bracket{
  \Psi^\dagger(t_y) ,  t V^*(y)  \Psi(t_y)} \sim i \bracket{
  \psi^\dagger_y ,  t   \pt_y^x V^*(x)\psi_x}$. 
This suggests an approximate heat kernel
 \be K_{\Delta^\XX}(x,y,\psi_x,\psi_y;t) = \bint H_D(x,y;t)
e^{-\Ricci\parens{\vec{x}_y,\vec{x}_y}/12 - t \scalar/12 + i
\bracket{\psi_y^\dagger,  \pt_y^x \sbrace{1-t V^*(x)/2}\psi_x -  \psi_y}}
d\psi_y^\dagger.
\ee{super-k}
Here
\be H_D(x,y;t)= \chi_{<D}(x,y)
  (2\pi t)^{-m/2} e^{-\abs{\vec{y}_x}^2/(2t)},
\ee{h-def}
where $\chi_{<D}$ provides a cut-off away from the diagonal
\[
\chi_{<D}(x,y) = \begin{cases}
	1& \quad \text{if} \quad d(x,y)<D\\
	0 &	\quad \text{else,}
\end{cases},
\]
and $D > 0$ is small enough that there is in fact a unique geodesic
between $x$ and $y$.  The kernel $H_D$, which for the Euclidean metric
agrees with the flat-space heat kernel for $d(x,y) < D$, will serve as
the basic kernel to which to compare all others.

The Ricci and scalar
curvature terms do not follow directly from the approximation to the
action. Rather, referring to Rem.~\ref{rm:correction}, they correspond to the resolution of the
operator-ordering ambiguity that gives $\Delta^\XX$ as the operator
whose kernel is the path integral with this Lagrangian, and, among
such choices, they are of the particular form to make
$K_{\Delta^\XX}$ an approximate heat kernel for $\Delta^\XX$ in the
technical sense required for the convergence arguments of
Sect.~\ref{ss:convergence} below. These ensure, 
under the tameness assumptions of Def.~\ref{def:tame},  the
time-slicing approximations to the path integral converge pointwise to the heat
kernel for $\Delta^\XX$.
\subsection{The Dirac operator \& twisted $N=1/2$ SUSYQM}
\subsubsection{The Dirac operator}
Heuristically, the path integral for twisted $N=1/2$ SUSYQM 
in imaginary time
is related to the kernel of the heat operator for a
Laplacian which is the square of the twisted Dirac
operator~\cite{FW84}. 
To define the twisted Dirac operator for a manifold,
 recall some Clifford algebra facts and terminology
 as detailed for instance in Ch. 3 of~\cite{BGV04}.
  If $M$ is a Riemannian manifold,
define $\Clifford=C(T^*M)$ to be the bundle which at each point $x\in M$  is the
complexified $\ZZ/2\ZZ$-graded (and $\ZZ$-filtrated) algebra generated by $T_x^*M,$ subject to the relation
\be
v^*\cdot  w^*+w^*\cdot v^*= -2\parens{v^*,w^*}.
\ee{clifford-def}
A \emph{Clifford module} is a graded  vector bundle $\VV$ over $M$
with a graded homomorphism
$c_\VV\colon\Clifford \to \End\parens{\VV}$.   $\Lambda\parens{T^*M}$ is a Clifford
module with the action $c_\Lambda(v^*)\alpha= v^* \wedge \alpha - i_v(\alpha)$
where $v$ is dual to $v^*$ in the inner product. 

If $M$ is even-dimensional and spin, the spinor
bundle $\spinor= \Lambda \PP,$ where $\PP$ is a polarization of the complexified cotangent bundle of
$M$ is  a Clifford module.
Indeed, with this action,  $\Clifford \iso \End\parens{\spinor},$ and any Clifford
module can be written as $\VV= \spinor \tensor \tspinor,$ where
$\tspinor$ is a vector bundle on which $\Clifford$ acts trivially.

For $\VV$  a Clifford module, a connection $\nabla^\VV$ is a \emph{Clifford
  connection} if, for any vector field $X$ and 
section $Y$ of $T^*M,$ 
\be
\sbrace{\nabla^\VV_X,c_\VV(Y)}=c_\VV\parens{\nabla_X^{\text{LC}}Y}.
\ee{cliffconn-def}
(The bracket on the left-hand side is graded.)
In the case where $M$ is even-dimensional and spin,  
any  Clifford connection  $\nabla^\VV$  can be written as 
\be
\nabla^\VV= \nabla^\spinor \tensor 1 + 1 \tensor \nabla^\tspinor
\ee{nabla-decomp}
for some connection $\nabla^\tspinor$ on $\tspinor$ and the Levi-Civita
connection $\nabla^\spinor$ on $\spinor$.  If $M$ is even-dimensional
but not spin, the Clifford action is still faithful and the 
curvature of a Clifford connection still decomposes as $R+F^\tspinor,$ 
where $R$ is  Riemannian curvature and $F^\tspinor$ is the component of the
curvature in $\End_{C(M)}(\VV)$. \cite{BGV04}(Props.
3.35,3.40  \& 3.43). 

If $\VV$ is a Clifford module and $\nabla^\VV$ a Clifford connection, the twisted  Dirac operator is
\be
\dirac^\VV= c_\VV(dx^i)\nabla^\VV_{\partial_i}.
\ee{dirac-def}
In the case of $R^n$ and trivial $\VV$, writing $c_\VV(dx^i) =
\gamma^i$ which acts on spinors of a given type, this is the standard
Dirac operator $\dirac = \gamma^i \partial_i$.
The square of $\dirac^\VV$ is a generalized Laplacian $\Delta^\VV$
with section $V =
c_\VV\parens{F^\tspinor} - \scalar/4$, where  $c_\VV$ acts on
two-forms by $c_\VV(v^* \wedge w^*) = \frac{1}{2}\sbrace{c_\VV(v^*)c_\VV( w^*) - c_\VV( w^*)c_\VV(v^*)}.$ 
That is, with this choice of $V$,
\be
\Delta^\VV = \parens{\dirac^\VV}^2.
\ee{dirac-squared}
In the special case $\VV = \spinor$, the operator $\dirac^\VV$ is
the ordinary Dirac operator.
\subsubsection{The Lagrangian for twisted $N=1/2$ SUSYQM and a
  time-slicing approximation to the corresponding path integral}
If $M$ is even-dimensional and spin 
and $\tspinor$ is a bundle over $M$ with a connection whose curvature is
$F$, define  \emph{twisted $N=1/2$ SUSYQM} 
via the
action
\be
S_{\twisted} = \int_0^t \frac{1}{2}\parens{\dot{\sigma},\dot{\sigma}} + i
 \bracket{\Psi^\dagger, \nabla^\spinor_s \Psi} + i
 \bracket{\Pi^\dagger, \nabla^\tspinor_s \Pi}- 
\frac{i}{2}\bracket{ F(\Psi, \Psi) \Pi^\dagger,\Pi} ds,
\ee{susyqm-action}
for $\Psi$ and $\Psi^\dagger$ Grassman-valued lifts of $\sigma$ to $\PP^*$ and $\PP$
respectively, 
and $\Pi$ and
$\Pi^\dagger$ Grassman-valued lifts to $\tspinor^*$ and $\tspinor$
respectively.
This action was first written down by Friedan and  Windey \cite{FW84}
(with slightly different normalization conventions). If $\tspinor$
is the trivial bundle it reduces to the action for 
$N=1/2$ SUSYQM of \cite{Alvarez83}.

Discretize as above to get a kernel
on $\hat{\VV}=\spinor \tensor \Lambda \tspinor$
\begin{align}
K_{\twisted} = &  \bint H_D(x,y;t)
e^{-\Ricci\parens{\vec{x}_y,\vec{x}_y}/12 - t \scalar/12}\label{eq:ksusy} \\
& \times e^{i\bracket{\psi_y^\dagger,  \pt_y^x \psi_x -  \psi_y} + i
\bracket{\eta_y^\dagger,  \pt_y^x \eta_x -  \eta_y} + i t 
\bracket{\eta_y^\dagger,  \pt_y^x\sbrace{F(\psi_x, \psi_x) +  \scalar/4} \eta_x}/2} 
d\eta_y^\dagger d\psi_y^\dagger,  \nonumber
\end{align}
where the parallel transports are with respect to the connections
$\nabla^\spinor$ and $\nabla^\tspinor$. As in the general case, the
terms with parallel transport represent, under Berezin integration,
the kernel of $e^{-\frac{1}{2} \sbrace{c(F) - \scalar/4}}
\pt^x_y$, with this parallel transport being with respect to the
connection 
on $\hat{\VV}$.
Thus the
discretization is exactly the approximate heat kernel
$K_{\Delta^{\hat{\VV}}}$ of Eq.~\eqref{eq:super-k} with the choice  $V = c(F) - \scalar/4$. 
\section{Convergence results for path integrals for generalized
  Laplacians}  \label{s:convergence_broad}
%
\subsection{Kernels, local geometry, and the $t$-norm}
%
As noted above, if the time-slicing approximate kernel happens to be the heat kernel
(as is the case for the free theory in Euclidean space) then the semigroup
property makes the convergence automatic. In a more general case, the
idea is to first show that the successive approximations of
Eq.~\eqref{eq:time_slicing} approach some limit as the partitions become
finer, and then to show that limit 
is the heat kernel. To see whether a limit exists, think of the effect of
subdividing one interval in a given partition. This replaces a term of
the form
$K(x, y; t)$ with $\int K(x, z; t_1) K(z, y; t_2) \,
dz$ where $t_1 + t_2 = t$. Writing $\int K(x, z; t_1) K(z, y; t_2) \,
dz = K(x, y; t) + \epsilon (x, y, t_1, t_2)$, where
$\epsilon$ denotes an error term reflecting the failure of $K$ to be a
semigroup, the question of convergence boils
down to keeping track of how the error terms propagate. That is, how
big are terms like
$\int \epsilon(x, z; t_1, t_2) K(z, y; t_3) \, dz$ and
$\int \epsilon(x, z; t_1, t_2) \epsilon(z, y, z; t_3, t_4) \, dz$?
Here ``big'' should mean as compared to the kernels  $K$ 
themselves. 

There are two issues: The obvious one is that as the
partitions become finer, the number of error terms, and of their products
with other kernels and each other, increases. The error terms
must decrease, in some measure of their size, quickly enough
as their time arguments decrease to ensure the
sum of the errors does not build
up to be infinitely large as the partitions become finer. The more subtle
issue is the kernels $K$ relative to which the error terms should be small
are actually families of operator kernels, with parameter
$t$, and that even the ``best'' example, the heat kernel, is singular
on the diagonal ($x = y$ in $K(x,y;t)$) as $t \to 0$. Thus, in asking whether
a given error term is ``big'', the comparison will be to something
that may be singular in places. Likewise, away from the diagonal the
heat kernel vanishes rapidly as $t \to 0$ or as the distance between
$x$ and $y$ increases, so ``small'' should be
in comparison to something with this rapid
decay. Sect.~\ref{ss:tnorm} defines a family $\EE'_{B, D}(t)$ of
kernels with this behavior, and a ``$t$-norm'' which takes these features into
account.

Placing an appropriate bound in the $t$-norm on the failure of a time-slicing
approximate kernel to be a semigroup suffices to prove the
convergence of the approximate path integrals under refinement. An additional
bound on the failure of the approximate kernel to satisfy the
heat equation will ensure that the limiting kernel is indeed the heat
kernel.\footnote{Without the heat equation bound, the cumulative effect of errors in the semigroup
property may not spoil convergence, but will in general allow the
limiting kernel to differ from the original time-slicing approximate
kernel; the bound ensures the limiting kernel is in fact the heat kernel.}
%
\subsubsection{Notation and some facts about local geometry in $R^n$ with
a non-Euclidean metric} \label{sss:local_geom}
%
The positive aspect of comparing error terms with kernels having the 
extreme behavior noted above is that the convergence arguments turn out to be
entirely local thanks to the rapid decay away from the diagonal. With this in mind, consider first an open set $O \in
R^n$ with smooth Riemannian metric $g$ (not necessarily Euclidean).
For technical reasons related to the convergence argument to follow,
require that all derivatives of order $k$ of $g$ and of $g^{-1}$ are
bounded in supremum norm for $0 \leq k \leq 5.$

Let $d(x,y)$ be the distance between $x,y\in O$ in this metric.  For
$\vec{v} \in \RR^m,$ $x \in O$ and $t \in \RR$ the geodesic through
$x$ with tangent $\vec{v}$ at $x$ and parameter $t$ proportional to
arc length defines the exponential map $\exp_x t \vec{v}$. If $y\in O$
is close enough to $x$ that there is a unique minimal geodesic
connecting them, define $\vec{y}_x=\exp^{-1}_x y;$ that is,
$\vec{y}_x$ is the vector at $x$ whose exponential gives
$y$.\footnote{While $d(x,y)$ is of course symmetric in the two points,
the notation here and in Eq.~\eqref{eq:h-def} suggests thinking of $x$ as fixed and $y$ as
variable, which is natural in the context of a kernel acting via
integration as in
Eq.~\eqref{eq:Kf}. Section~\ref{ss:convergence}, which applies
the Laplacian to specific kernels, reverses this to allow the operator
to act on the first variable, as is natural in this context. The
switch is purely a matter of convention.} Let
$\parens{\,\cdot,\cdot\,}_x$ denote the inner product with respect to
$g$ at $x \in O,$ and let $\abs{\,\cdot\,}_x$ denote the corresponding
norm. If the vectors inside the norm or inner product are of the form $\vec{y}_x$, or the point
at which this  is computed is otherwise understood
from context, drop the subscript.  Write $\dmeas_g
y= \det^{1/2}_y(g)\dmeas y,$ where $\dmeas y$ is standard Lebesgue
measure on $\RR^m$ restricted to $O$, and write $\dmeas
\vec{y}_x$ for Lebesgue measure on $O$ with respect to the inner
product given by $g$ at $x;$ that is, the metric measure at $x$ pulled
back to $y$ by $\exp^{-1}_x$.

Henceforth to say that a quantity, such as $D$ in the following lemma,
``depends on the metric bounds'' will mean that quantity is a
function of the assumed bounds on the supremum norm of
 $g$, $g^{-1}$
and their first five derivatives (as well as on the dimension
$m$).
 The concern is that, in later
arguments which require rescaling the metric, preserving these bounds
should be
sufficient to preserve the estimates which follow here.

Even with the
general metric, nearby points in $O$ behave a lot like points in
Euclidean space, as regards length and integration.
Specifically, direct arguments based on Rauch's comparison
theorem~\cite{doCarmo92} show there is 
 is a $D>0$ depending on the metric bounds  such that,
        for $x,y,z\in O$ with $d(x,y),d(y,z),d(x,z)<D$,  there is a unique minimal
        geodesic connecting $x$ and $y$,  $\vec{y}_x$
	depends smoothly on $x$ and $y$,  
        and $y-x$ depends smoothly on $x$ and on $\vec{y}_x$.
Moreover, 
	\begin{align}
          y-x&= \vec{y}_x + \OO\parens{\abs{\vec{y}_x}^2} \label{eq:yminusx-est}\\
          \abs{\vec{z}_x}^2 &= \abs{\vec{z}_y}^2 +\abs{\vec{x}_y}^2 -
          2 \parens{\vec{x}_y,\vec{z}_y} +
          \OO\parens{\abs{\vec{x}_y}^2
            \abs{\vec{z}_y}^2}\label{eq:lengthsquared}\\ 
         \frac{\dmeas_gy}{\dmeas \vec{y}_x}&=1+\OO\parens{\abs{\vec{y}_x}^2} \label{eq:determinant}
	\end{align}
	where for example $\OO\parens{\abs{\vec{x}_y}^2
          \abs{\vec{z}_y}^2}$ 
        indicates the difference between the left-hand side and the
        truncated Taylor series is bounded by a constant (depending on
        the metric bounds) times 
        $\abs{\vec{x}_y}^2 \abs{\vec{z}_y}^2$
        (as each of these tends towards zero).
%
\subsubsection{The operator norm, and
        the ``kernel'' norm}
%
Given $n \in \NN,$ let $f\colon O \to \RR^n,$ $f^* \colon O
\to \parens{\RR^n}^*$ and $K \colon O \times O \to
\operatorname{Matrix}_{n,n}$. Notice $f$ and $f^*$, as functions from
$O$ to $\RR^n$ or $\parens{\RR^n}^*$, are local expressions of
sections of vector bundles, and $K$ represents kernels of the left or
right operators (on the space of such functions) whose actions are given
by
\eqa{*-def}{
	K*f(x)&= \int_O K(x,y) \cdot f(y) \dmeas_gy \nonumber\\
	f^**K(y)&= \int_O f^*(x) \cdot K(x,y) \dmeas_gx 
}
where $\cdot$
represents the matrix product. The kernel of the operator product of
the operators represented by $K$ and $J$ is the $*$-product 
\eqa{*-product}{
J*K(x,z)&= \int_O J(x,y) \cdot K(y,z) \dmeas_gy.
}

The matrix norm sends $K$ to a nonnegative function $\abs{K}$ on
 $O \times O.$ Use this to define
\[
\norm{K}_{\mathrm{op}}= \max\parens{\sup_x \int \abs{K(x,y)} \dmeas_g
 y, \sup_y \int  \abs{K(x,y)}\dmeas_g x},
 \]
 which is the maximum of the operator norms of $K$ acting on the left and
 the right. 
 Define the kernel norm by
 \[
 \norm{K}_\mathrm{ker}=\max(\norm{K}_\mathrm{op},\norm{K}_\infty).
 \]
 Notice $\norm{J*K}_\mathrm{ker} \leq \norm{J}_\mathrm{ker}\norm{K}_\mathrm{ker}$ and $\norm{J*K}_\mathrm{ker} \leq
 \norm{J}_\mathrm{op}\norm{K}_\mathrm{ker}$.
%
\subsubsection{Two families of kernels and the $t$-norm} \label{ss:tnorm}
%
Now begin to explore classes of kernels whose relation to $H_D$
are increasingly tenuous, to delineate the extent to which they retain
key properties of the heat kernel under kernel products. This
exploration culminates in the definition of a class of kernels
$\EE'_{B, D}(t)$ against which to compare
 error terms like those above, as well as others arising from the failure of the time-slicing
approximation to satisfy the heat equation.  Appropriate bounds on
these errors, expressed in terms of the ``$t$-norm'', which measures
the ratio of the error to elements of $\EE'_{B, D}(t)$, will ensure the approximate
path integrals converge with sufficient rapidity to the heat kernel of
the given Laplacian.
For $B$ large enough,  $D$ small enough, and $t$  small
        enough for
        the right-hand side to make sense (each depending 
        on the bounds of the metric and the preceding quantities), define
       \label{lm:h*h-bd}
        \be
        K_{B,D}(x,y;t)=e^{B\abs{\vec{y}_x}^2/(5m)}H_{D}(x,y;t),
        \ee{kbd-def}
        This allows $K_{B, D}$ to grow much faster than $H_{D}$ away
        from the diagonal (for fixed $t$). Nevertheless,
        for      $0<t_1,t_2,$ and $t=t_1+t_2,$ 
 	\eqa{h*h-bd}{
 	  \chi_{<D}\sbrace{K_{B,D}(t_1)*K_{B,D}(t_2)} &\leq  e^{Bt_1t_2/t}K_{B,D}(t),\nonumber \\
 	\norm{\chi_{>D}\sbrace{K_{B,D}(t_1)*K_{B,D}(t_2)}}_{\mathrm{ker}} & \leq t^2e^{-D^2/9t},
 	}
 	which gives control over kernel products both near and far
        from the diagonal. Moreover, as an operator, 
 	\be
 	\norm{K_{B,D}(t)}_{\mathrm{op}}\leq e^{Bt}.
 	\ee{hopnorm-bd}
 The derivation of these inequalities
 uses the facts about local geometry from Eqs.~\ref{eq:lengthsquared}
 and~\ref{eq:determinant} to bound
 the Gaussian integrals implicit in Eqs.~\ref{eq:h*h-bd} and~\ref{eq:hopnorm-bd}.

Now smear out $K_{B, D}$ a little in time and allow some additional
growth in $t$ to define a class of kernels $\EE_{B, D}(t)$ which will
be almost closed under the $*$ product and whose products, away from
the diagonal, decay 
 rapidly with decreasing $t$, in the kernel norm, as do those of the heat
kernel. 
\begin{definition} \label{def:EE}
For $B,D,t>0$ define $\EE_{B,D}(t)$  to
be the set of all kernels $K$ 
for which
there exists a probability measure $\dmeas\mu$ on the interval $[1,2]$
such that 
\be
         \abs{K(x,y)}  \leq e^{B\sqrt{t}}\int K_{B,D}(x,y;\alpha t)
        \dmeas \mu_\alpha,
\ee{ebd-def}
where $K_{B,D}$ is the particular one-parameter family of kernels defined in
Eq.~\eqref{eq:kbd-def}.
\end{definition}
Note that $K_{B,D}(t)$ 
itself is in
$\EE_{B,D}(t)$. Direct estimates using the above properties of $K_{B,
D}$ lead to the following precise statements about this class:
	If $B$ is large enough, $D$ is small enough, and $T$ is small
        enough (each depending  on
        the bounds of the metric and the previous quantities) and if
        $K_1$ and $K_2$ are one-parameter families of kernels with
        $K_1(t),K_2 (t) \in \EE_{B,D}(t)$  for $t < T,$ then, for
        $0<t_1,t_2$ and $t=t_1+t_2 < T$ 
	\be
	\norm{K_i(t)}_{\mathrm{op}}\leq e^{1.1 B\sqrt{t}}
	\ee{kopnorm-bd}
	and
	\eqa{k*k-bd}{
	  \chi_{<D}K_1(t_1)*K_2(t_2) &\in e^{B\sqrt{t_1t_2/t}}\EE_{B,D}(t)\nonumber \\
	\norm{\chi_{>D}K_1(t_1)*K_2(t_2)}_{\mathrm{ker}} & \leq t^2e^{-D^2/(20t)}.
	}

Continue to enlarge the class of kernels which behave  well
under kernel products to allow an additional ``tail'' behavior for
larger $t$ to get the final class of kernels against which to measure
various error terms: 
\begin{definition} \label{def:EE'}
For, $B,D,t>0$ define $\EE'_{B,D}(t)$ to be the set
of all kernels which can be written as $K  + J$ where $K \in
\EE_{B,D}(t)$ and $\norm{J}_\mathrm{ker}\leq te^{-D^2/(20t)}$. 
\end{definition}

The advantage of incorporating the tail behavior into the class of
kernels is products of kernels in this class almost stay within the
class. The precise statement of the properties of kernels in $\EE'_{B,
D}(t)$ 
follow easily from the definitions:
If $B$ is large enough,  $D$ is small enough and $T$ is small
        enough (each depending only
        on the bounds of the metric and the previous quantities)
        and if 
        $K_1$ and $K_2$ are one-parameter families of kernels with
        $K_1(t),K_2 (t) \in \EE'_{B,D}(t)$  for all $t < T,$ then, for
        $0<t_1,t_2$ and $t=t_1+t_2 < T$ 
 	\be
	\norm{K_i(t)}_{\mathrm{op}}\leq e^{2B\sqrt{t}},
	\ee{kop-bd} 
 
	\be
	\abs{K_i(x,y;t)} \leq 2 (2 \pi t)^{-m/2} e^{-d(x,y)^2/(4t)} + t e^{-D^2/(20t)},
	\ee{inf-bd}
    and    	
        	\be
	  K_1(t_1)*K_2(t_2) \in e^{B\sqrt{t}}\EE'_{B,D}(t).
	\ee{*-bd}

The class $\EE'_{B, D}(t)$ and its properties explicitly depend on
	choices of constants 
	$B$, $D$ and  $T$ (the last as an upper bound for $t$). The relation of
	these  constants to the bounds on the metric and the relation
	between these constants are as follows:
	There is a minimum $B$ and a maximum $D$ and $T$  to make
        the above properties hold, and these numbers depend only
        on the supremum of the first few derivatives of the metric and
        its inverse (and on the dimension $m$), a fact that will be crucial in
        Sect.~\ref{sss:rescale}.
       Choosing  a larger $B$ would make the 
        maximum $D$ and $T$ smaller, but these would still exist.
        If one chose an even smaller $D,$ the maximum $T$ would be
        smaller still.   In the definition of approximate
        semigroup and approximate heat kernel below, the choice of
        constants will further depend on the family of kernels being
        considered.

The properties of $\EE'_{B, D}(t)$ provide the basis upon which to
define a norm, which indeed is the motivation for defining this class:
\begin{definition} \label{def:t-norm}
For given $B, D, t > 0$  define the $t$-norm $\norm{K}_{(t)}$ to be the
smallest positive real number such that $K/\norm{K}_\tnm{t} \in
\EE'_{B,D}(t)$ if it exists. (Otherwise set $\norm{K}_\tnm{t} =
\infty$.) 
\end{definition}
The advantage of using this norm is its behavior under the kernel
product:  If $B$ is large enough,  $D$ is small enough and $t$ is small enough
  (each depending only on the bounds of the metric and the previous
  constants), then  for the associated
        $t$-norm and for families of kernels $K_1,$ and $K_2$,
\be
   \norm{K_1(t_1)*K_2(t_2)}_\tnm{t}\leq
        e^{B\sqrt{t}}\norm{K_1}_\tnm{t_1}\norm{K_2}_\tnm{t_2}.
\ee{t*-bd}
At the same time, the new norm is related to the more obvious norms
via
	\be
	\norm{K_i}_\mathrm{op} \leq e^{2B\sqrt{t}} \norm{K_i}_{\tnm{t}},
	\ee{top-bd}
 and
	\be
	\abs{K_i(x,y;t)} \leq  \norm{K_i}_\tnm{t}\sbrace{2 (2 \pi t)^{-m/2} e^{-d(x,y)^2/(4t)} + t e^{-D^2/(20t)}};
	\ee{tpntws-bd}
 In particular, there is an $A_2>0$ such that 
	\be
	\norm{K_i(t)}_\infty \leq  A_2t^{-m/2}\norm{K_i}_\tnm{t}.
	\ee{tinf-bd}
The above are all restatements or immediate consequences of the
properties of kernels in the class
$\EE'_{B, D}(t)$. 

\subsection{Approximate semigroups \& approximate heat kernels}
\subsubsection{Approximate semigroups on $O \subset R^m$}
With the $t$-norm in hand, define an \emph{approximate semigroup} with
constants $(B,C,D,T)$ as a family of kernels $K(t)$ for which
\be
	 \norm{K(t)}_{(t)}\leq 1
\ee{k-bd}
and, given $0<t_1,t_2$ and $t=t_1+t_2<T$,
\be
	\norm{K(t_1)*K(t_2)-K(t)}_{(t)}\leq Ct^{3/2}.
\ee{k*k-est}
Notice Eq.~\eqref{eq:k-bd} means an approximate semigroup must be in the
class $\EE'_{B, D}(t)$  for all $t<T$. Some easy estimates show the
 condition keeping $K*K$ close to $K$ reduces to a condition only 
on the piece  $\widetilde{K}(t) \in \EE(t)$ in the decomposition
$K(t)=\widetilde{K}(t) + J(t)$.

\subsubsection{Approximate heat kernel on $O \subset R^m$}
Now consider kernels in $\EE'(t)$  with additional conditions tying
them to a generalized Laplacian. Specifically, given a generalized
Laplacian $\Delta$ define  an \emph{approximate heat kernel}
for $\Delta$ with constants $(B,C,D,T)$, all positive, as 
a family of kernels $K(t)$ whose members are differentiable to first
order in $t \in (0,T)$ and to second order in the 
  spatial variables, and satisfy, for $t<T$ and the $t$-norm with constants $(B,D),$
  \be
  \norm{K(t)}_{(t)}\leq 1,
  \ee{k2-bd}
  for all $f\colon O \to \RR^n$
  \be
  \lim_{t \to 0} K(t)*f =f, 	
  \ee{tto0}
  \be 
  \lim_{t \to 0} \frac{K(t)*f -f}{t}  = \frac{\Delta}{2} f 
  \ee{kf}
  (both pointwise),
  \be
  \norm{\frac{\partial}{\partial x}K(x,y;t)}_{(t)},\norm{\frac{\partial}{\partial y}K(x,y;t)}_{(t)}\leq B/t,
  \ee{kprime-bd}
  and
  \eqa{ahk}{
    \norm{\parens{\frac{1}{2}\Delta_x  - \frac{\partial}{\partial t}}
      K(x,y;t)}_\tnm{t} & \leq  Ct^{1/2}\nonumber \\
    \norm{\parens{\frac{1}{2}\Delta^*_y  - \frac{\partial}{\partial
          t}} K(x,y;t)}_\tnm{t} & \leq Ct^{1/2},
  }
  where
  $\Delta_x$ acts from the left on  $\End(\RR^n)$ and  $\Delta_y^*$ acts
  from
  the right via $\int_{O} \Delta_y^*[h^*(y)] \cdot f(y)
  \dmeas_g y = \int_{O} h^*(y) \cdot \Delta_y[f(y)]
  \dmeas_gy.$
  The first condition again ensure $K(t)$ is in $\EE'(t)$. The next
  ensures the operator $K$ defines will implement the 
  initial condition expected of the heat operator. Eq.~\eqref{eq:kf}
  says this operator agrees with the heat operator as $t \to 0$;
  whereas, Eq~\eqref{eq:ahk} bounds the failure of $K$ to satisfy the
  heat equation (which the heat kernel would) away from $t=0$.  The bound of
  Eq.~\eqref{eq:kprime-bd}, 
  along with those of Eq.~\eqref{eq:ahk} and the observation regarding
  checking Eq.~\eqref{eq:k*k-est}, combine with straight-forward estimates
  to show an approximate heat kernel in the sense of this definition
  satisfies Eq.~\eqref{eq:k*k-est} and is thus also an approximate
  semigroup. (The constants $C$, $D$ and $T$ may need to be refined in
  passing from the approximate heat kernel to the approximate
  semigroup.)
\subsubsection{Approximate semigroups and approximate heat kernels on
tame manifolds}
  Recall that so far everything has taken place on an open set $O
  \subset R^m$. To define the $t$-norm for kernels on the tame
  manifolds of
   Sec.~\ref{sec:tame}, simply observe that
  on any tame atlas, for sufficiently large $B$ and sufficiently
small $D$, there is a sufficiently small $t$ such that
the $t$-norm with constants $(B,D)$ can be defined on each chart.
Define $\norm{K}_{\tnm{t}}$ to be the
supremum of the $t$-norms of its image in each chart.
If $(M,g,\VV)$ is tame the $t$-norm defined in terms of any tame
  atlas will satisfy Eqs.~\eqref{eq:t*-bd} and~\eqref{eq:top-bd}
  for
  sufficiently large $B$ and sufficiently small $D$.
Then extend the definition of an approximate semigroup to be a family
  of kernels $K(t)$ on $\VV$ for which $(M,g,\VV)$
admits a tame atlas on each chart of which $K$ is  represented as an
approximate semigroup. Require of course that the constant $D$
implicit in the definition of an approximate semigroup on the chart be
less than the constant $D_0$ appearing in the definition of a tame
atlas.  Extend the definition of approximate heat kernel
analogously. The bounds of Eqs.~\eqref{eq:t*-bd}-\eqref{eq:k*k-est}
  extend to
any approximate semigroup on $\VV$.  Again because all the previous
results were local, an approximate heat kernel 
for some $\Delta$ on $\VV$ is an approximate semigroup. The
constants $(B,C,D,T)$ of this approximate semigroup can be taken to depend
only on the corresponding constants for the approximate heat kernel
and the bounds on the defining atlas.

\begin{remark}
  While it suffices for the rest of the work, the dependence of the
  structures defined on the choice of tame atlas might distress the  mathematically
 inclined reader.  However,  there is a natural notion of the comparability
  of tame structures, which simply involves requiring that the
  diffeomorphisms between charts induced by the identity on $\VV$ have
  all derivatives up to the appropriate order uniformly bounded.  It
  is then straightforward if laborious to check that the $t$-norms
  associated to compatible tame atlases are comparable (each bounded
  by a multiple of the other), that families of kernels that are
  approximate semigroups or heat kernels with respect to one atlas
  are the same with respect to the other, and therefore that the limit
  results of the following section depend only on the ``tame
  equivalence class'' of the vector bundle, Riemannian manifold and
  operator.
\end{remark}
\subsubsection{The twisted $N=1/2$ SUSYQM time-slicing approximation as
  an approximate heat kernel}
Return at last to the time-slicing approximate heat kernel
$K_{\Delta^\XX}$ of Eq.~\eqref{eq:super-k} to check it is in fact an
approximate heat kernel in the specific sense of the preceding
definitions. First use standard properties of integration with
Grassman variables to see the quantity
\[\bint e^{i\bracket{\psi^\dagger_y, \pt^x_y \sbrace{1-t V^*(x)/2}\psi_x-  \psi_y}}
d\psi_y^\dagger\]
is, up to terms in $\OO\parens{t^2}$,  the superkernel for the
 operator
\[e^{-tV(x)/2} \pt^y_x \colon \XX_y \to \XX_x,\] which is
the 
extension 
of $e^{-tV(x)/2} \pt^y_x : \VV_y \to \VV_x$. As the addition of terms of order
$\OO\parens{t^2} K$ will affect neither whether a kernel $K$ is an
approximate heat kernel, nor convergence of kernel products nor the
fine-partition limit, consider the kernel\footnote{Up to the
above-mentioned irrelevant terms, $K_\Delta$ is just $K_{\Delta^\XX}$
of Eq.~\eqref{eq:super-k}, but
written in a more general form that would describe a kernel on a
bundle $\XX$ that need not be of the form $\Lambda \VV$.}
\be
K_\Delta(x,y;t)= H_D(x,y;t)e^{ -\Ricci\parens{\vec{x}_y,
            \vec{x}_y}/12 - \scalar t/12 - tV(x)/2} \pt^y_x.
\ee{kgen-def}
Checking that $K_\Delta$ is an approximate heat kernel means checking
it satisfies equations Eqs.~\eqref{eq:k2-bd} through~\eqref{eq:ahk}. 
These
  are all local conditions, so  pick $y \in M$ and work  in Riemann
  normal coordinates centered at $y.$ 
That is, choose an orthonormal basis for $T_{y}M$,  and notice each point $x \in M$ near $y$ is the value of the exponential map at a
unique vector $\vec{x} \in T_{y}M$ near $0$. (The $\vec{x}$ 
 was $\vec{x}_{y}$ in Sect.~\ref{sss:local_geom}; the subscript is implicit here where there is no
danger of confusion.) The components of $\vec{x}$ with respect to the
chosen basis define the Riemann normal coordinates of the point $x$. 
Some thought about the the geodesic equation as a system of ODE's,
which is also the basis of the facts laid out in Sect.~\ref{sss:local_geom},
says  tameness implies  $g_{ij}$ in Riemann normal
coordinates  has bounded $k$th derivatives for $0 \leq k \leq 4.$

If $X$ and $Y$ are tangent vectors at $x \in M$ let $R_x[X,Y]$ be the
Riemannian curvature (endomorphism on $T_xM$), $\Ricci_x(X,Y)$ be the
Ricci curvature, and $\scalar_x$ be the scalar curvature.
The coordinate derivatives $\partial_i$ for $i=1, \ldots,m$  at
each $x \in M$ near $y \in M$  form
 a basis of $T_{x}M$ and define
vector fields in a neighborhood of $y$ 
(commuting but not in general orthonormal). At $y$ these agree with
the original choice of orthonormal basis. Define a second basis
$e_i \in T_{x}M$ (orthonormal but not commuting as vector fields) 
by parallel transporting the same orthonormal basis of
$T_{y}M$ along a minimal geodesic from $y$ to (nearby) $x$.  
The two bases are related by \cite{BGV04}(Prop. 1.28)
\be e_i = \sbrace{\delta_i^j + \frac{1}{6}R_{ikl}^{\rule{1.2em}{0em}j} x^k
  x^l}\partial_j + \OO\parens{\abs{\vec{x}}^3}  
\ee{e-partial} 
where
$R_{ikl}^{\rule{1.2em}{0em}j}\partial_j=R_y[\partial_i,\partial_k]\partial_l$
defines 
the coordinates of the curvature at $y.$
If $g_{ij}\parens{\vec{x}}=
\parens{\partial_i, \partial_j}_x,$
 with
inverse $g^{ij}\parens{\vec{x}},$ and 
 $
  \Gamma_{ij}^k\parens{\vec{x}}\partial_k=
    \nabla_{\partial_i}^{\text{LC}} \partial_j,$
    Eq.~\eqref{eq:e-partial} implies 
\begin{align}
g_{ij}\parens{\vec{x}}&= \delta_{ij} + \frac{1}{3} R_{ikjl} x^k x^l
+ \OO\parens{\abs{\vec{x}}^3} \label{eq:g-rnest}\\
g^{ij}\parens{\vec{x}}&= \delta^{ij} - \frac{1}{3}
R_{k \rule{.2em}{0em} l}^{\rule{.3em}{0em} i \rule{.2em}{0em} j}  x^k x^l 
+ \OO\parens{\abs{\vec{x}}^3} \label{eq:ginv-rnest}\\
\Gamma_{ij}^k\parens{\vec{x}}&= - \frac{1}{3}
\sbrace{R_{ilj}^{ \rule{.8em}{0em} k} + R_{jli}^{\rule{.8em}{0em} k}}x^l
+ \OO\parens{\abs{\vec{x}}^2} \label{eq:gamma-rnest}\\
{\det}^{1/2}g(\vec{x})&= 1 + \frac{1}{6}
R_{ikj}^{\rule{1.1em}{0em}k}x^i x^j +
\OO\parens{\abs{\vec{x}}^3} \label{eq:dmeas-rnest}
 \end{align}
 freely raising and lowering indices using 
$g_{ij}(0)=\delta_{ij}.$  
At $y, $ abbreviate $\Ricci_y\parens{\partial_i,\partial_j}$ as 
$R_{ikj}^{\rule{1.1em}{0em} k}=\Ricci_{ij}$ and $\scalar_y$ as 
$\Ricci_i^i=\scalar.$ 

The bounds implicit in $\OO\parens{\abs{\vec{x}}^p}$ above depend only
on the
bounds on $g_{ij}$ and its derivatives up to order three. 
Trivialize
the bundle $\XX$ in a ball of radius $D$ around  $y$ by identifying
$\XX_x$ with $\XX_y$ via parallel transport along the unique minimal
geodesic connecting $y$ and $x.$ Then as in Prop.~1.8 of \cite{BGV04}
\be \nabla_i^\XX= \partial_i + \frac{1}{2}x^jF_{ij}^\XX + \OO\parens{\abs{\vec{x}}^2}
\ee{nabla-rnest}
where $F_{ij}$ is the curvature of $\nabla^\XX$ evaluated at $y$ in
the $\partial_i \wedge \partial_j$ direction, and the bound depends on
the bound on the 
coefficients of $\nabla$ to order $2.$ 

From its definition in Eq.~~\eqref{eq:kgen-def}, $K_\Delta(x,y; t)
= \sbrace{1 +
        \OO\parens{\abs{\vec{x}}} + \OO\parens{t}}H_D(x,y;t)$
Use the estimate
\be 
    d(x,y)^k H_D(x,y;t) \leq 2^{(m+k)/2}(k/e)^{k/2}{t^{k/2}} H_D(x,y;2t),
    \ee{time-space}
which follows  readily for $k \in \NN$ from  $x^k e^{-x^2/2} \leq
    (k/e)^{k/2}$,  to convert the $\vec{x}$ dependence to $t^{1/2}$ at the
    expense of doubling the time, giving
\[
K_\Delta(x,y; t)H_D(x,y;t) = H_D(x,y;t) + \OO\parens{t^{1/2}}H_D(x,y;2t).
\]
Then, for sufficiently large $B$, small enough $t$ and an appropriate $\mu$,
\begin{align*}
\abs{K_\Delta(t)}& \leq H_D(t) + \parens{e^{Bt^{1/2}}-1}H_D(2t) \\
&\leq e^{Bt^{1/2}}\sbrace{e^{-Bt^{1/2}} H_D(t)
  + \parens{1-e^{-Bt^{1/2}}}H_D(2t)} = e^{Bt^{1/2}}\int_1^2 H_D(\alpha
  t)d\mu_\alpha \in \EE_{B,D}(t)
\end{align*}
by
Def.~\ref{def:EE}.
Thus, 
$\norm{K_\Delta(t)}_{\tnm{t}}\leq 1,$ verifying  Eq.~\eqref{eq:k2-bd}
of the definition of an approximate heat kernel. 

Using Eqs.~\eqref{eq:g-rnest}-\eqref{eq:nabla-rnest} and the
antisymmetry of $F^\XX,$
\begin{align*}
\Delta&=g^{ij} \sbrace{\nabla_{i}^\XX \nabla_{j}^\XX
  - \Gamma_{ij}^k\nabla_{k}^\XX} - V\\
&=g^{ij}\sbrace{\partial_i \partial_j + \frac{1}{2} F^\XX_{ji}+
  \frac{1}{2}x^k\parens{F^\XX_{ik}\partial_j+F^\XX_{jk}\partial_i} -
  \Gamma_{ij}^k \partial_k +\OO\parens{\abs{\vec{x}}} +\OO\parens{\abs{\vec{x}}^2} \partial_i} - V\\ 
&= \partial^i \partial_i -\frac{1}{3}R_{k \rule{.2em}{0em} l}^{\rule{.3em}{0em} i \rule{.2em}{0em} j} 
x^k x^l\partial_i \partial_j - x^k F^{\XX,i}_{k}\partial_i -V
+\frac{2}{3} \Ricci_i^jx^i\partial_j +  \OO\parens{\abs{\vec{x}}} +
\OO\parens{\abs{\vec{x}}^2} \partial_i +
\OO\parens{\abs{\vec{x}}^3}\partial_i \partial_j. 
\end{align*}

Compute
\[\frac{\partial }{\partial t} K_\Delta(x,y;t) = \sbrace{ -\frac{m}{2t} + \frac{\abs{\vec{x}}^2}{2t^2} - \frac{\scalar}{12} - \frac{V}{2}}K_\Delta(x,y;t),\]
\[\partial_{i,x} K_\Delta(x,y;t) = \sbrace{-\frac{x_i}{t} -\frac{\Ricci_{ij}x^j}{6} + \OO(t)}K_\Delta(x,y;t),\] 
\[\partial_{i,x} \partial_{j,x}K_\Delta(x,y;t) =
\sbrace{-\frac{\delta_{ij}}{t} - \frac{\Ricci_{ij}}{6}+ \frac{x_i
    x_j}{t^2} +\frac{x^i\Ricci_{jk}x^k+x^j\Ricci_{ik}x^k }{6t} + \OO(t
  + \abs{\vec{x}}^2)}K_\Delta(x,y;t),\]  
so
\begin{align}
  & \sbrace{\frac{\partial}{\partial t} - \frac{1}{2}\Delta}K_\Delta
  =\Big[  -\frac{m}{2t} + \frac{\abs{\vec{x}}^2}{2t^2} -
  \frac{\scalar}{12} - \frac{V}{2}  + \frac{m}{2t} +
  \frac{\scalar}{12} -\frac{\abs{\vec{x}_y}^2}{2t^2} -
  \frac{x^i\Ricci_{ij}x^j}{6t} \\ 
&\qquad  - \frac{x^i\Ricci_{ij} x^j}{6t} + \frac{1}{6t^2}R_{kilj} x^k x^lx^i
x^j + \frac{1}{2t} x^kF^\XX_{ik} x^i + \frac{V}{2} + \frac{x^i\Ricci_{ij}x^j}{3t}\\
&\qquad + \OO\parens{\abs{\vec{x}}+\abs{\vec{x}}^3/t +
  \abs{\vec{x}}^5/t^2+ t}\Big]K_\Delta(x,y;t) \\ 
&=\OO\parens{\abs{\vec{x}}+\abs{\vec{x}}^3/t + \abs{\vec{x}}^5/t^2+
  t}K_\Delta(x,y;t) \label{eq:heat_errors}
\end{align}
after taking into account the antisymmetry of $F^\XX$ and the fourfold
symmetry of $R.$ Again using the estimate of  Eq.~\eqref{eq:time-space}, the right-hand side has
$t$-norm bounded by a multiple of $t^{1/2},$ verifying the first line of
Eq.~\eqref{eq:ahk}.  Since the Laplace-Beltrami operator is
self-adjoint, $\Delta^*$ is the operator associated to $g,$
$\nabla^\dagger$ and $V^\dagger,$ where $\dagger$ represents the
canonical map sending $\End(\RR^n)$ to $\End\parens{(\RR^n)^*}.$ So
for the second line of Eq.~\eqref{eq:ahk} it suffices to observe that
$K_{\Delta^*}(x,y;t)= K_\Delta^\dagger(y,x;t) + \OO(\abs{\vec{x}_y}^3
+ \abs{\vec{x}_y}t)$.
This estimate follows from the tameness
assumption which more directly implies
$\Ricci_x(\vec{y}_x,\vec{y}_x)- \Ricci_y(\vec{x}_y,\vec{x}_y)=
\OO\parens{\abs{\vec{x}_y}^3},$ $\scalar_x - \scalar_y =
\OO\parens{\abs{\vec{x}_y}},$ and
$V(y) - \parens{\pt_x^y}^{-1}V(x) \pt_x^y= \OO\parens{\abs{\vec{x}_y}}$,
with the bounds depending on the bounds on the
metric. Eq.~\eqref{eq:ahk} now follows.

For Eq.~\eqref{eq:tto0}, let $f$ be a smooth function on $O$ valued
in $\RR^n.$   Then, working in Riemann normal coordinates around $x$
with the the bundle trivialized by parallel transport in radial
directions,  
\begin{align*}
	\lim_{t \to 0} \int K_{\Delta}(x,y;t) \cdot f(y) \dmeas y &=
        \int H_D(x,y;t) f(y) \sbrace{1 + \OO\parens{\abs{\vec{y}_x}^2}
          + \OO(t)} \dmeas \vec{y}_x\\ 
&= f(x) + \OO(t).
\end{align*}
Similarly, for Eq.~\eqref{eq:kf} it suffices by 
the Mean Value Theorem 
to show $\lim_{t \to 0}\frac{\partial}{\partial
  t}K_\Delta*f = \frac{1}{2}\Delta f.$ In Riemann normal coordinates
\begin{align*}
\lim_{t \to 0}\frac{\partial}{\partial t} K_\Delta*f (x) &= \lim_{t \to
  0}\int \sbrace{ -\frac{m}{2t} + \frac{\abs{\vec{x}_y}^2}{2t^2} -
  \frac{\scalar}{12} - \frac{V}{2}}K_\Delta(x,y;t) f(y)\dmeas y\\ 
&= \lim_{t \to 0}\frac{1}{2}\sbrace{\partial_i \partial_i -V}f(x) + \OO\parens{t^{1/2}} = \frac{1}{2} \Delta f
\end{align*}
by straightforward Gaussian integrals.  Finally
Eq.~\eqref{eq:kprime-bd} follows for appropriate $B$ from the above
calculation for $\partial_i K_\Delta.$

\begin{remark} \label{rm:correction} The calculations verifying Eq.~\eqref{eq:ahk} shed some
  light on the role of 
  the  Ricci and scalar curvature terms in the definition of
  $K_\Delta.$ Absent the Ricci term and scalar terms, Eq.~\eqref{eq:heat_errors} would
  gain a net $\frac{1}{6t}\Ricci\parens{\vec{x}_y,\vec{x}_y}$, coming
  from the expansion of the metric and Christoffel symbols, which would persist as an
  $\OO(1)$ term in the $t$-norm, so Eq.~\eqref{eq:ahk} would fail to
  hold. Adding in just the Ricci term would cancel this; however, it would
  introduce an extra $\frac{\scalar}{12}$ to
  Eq.~\eqref{eq:heat_errors}, again an $\OO(1)$ term in the $t$-norm,
  and again spoiling  Eq.~\eqref{eq:ahk}. This scalar curvature term
  appears due to a general phenomenon, familiar from purely quadratic
  path integrals in $R^n$, where, for any self-adjoint linear operator
  $A$, $\int e^{-\frac{1}{2} \int_0^t
  |\sigmadot|^2 \, ds} \sbrace{\int_0^t (\sigmadot, A \sigmadot) \, ds} \, \DD \sigma= \tr A$. In
  more generality, a tedious calculation checks that, for an approximate heat kernel $K$, the
  modified kernel $K_A = K (1 + (\vec{x}, A \vec{x}) - t \tr A)$ has the
  property that $K_A^{*n} = K^{*n}\parens{1
  + \epsilon \sbrace{(\vec{x}, A \vec{x}) - t \tr A}}$, where
  $\epsilon$ depends on the precise nature of the partition, but,
  under mild restrictions, vanishes as the partition is  refined. 

The
  combination appearing in $K_\Delta$ thus cancels out the Ricci
  curvature terms, without adding new scalar terms. 
More generally, adding a term of the form $a\scalar t +
b \parens{\Ricci\parens{\vec{x}_y,\vec{x}_y}-\scalar t}$ to the
  exponent in  $K_\Delta$ adds error terms 
  $a\scalar + 2\norm{b \parens{\Ricci\parens{\vec{x}_y,\vec{x}_y}/t-\scalar}}_{(t)}
  \in \OO\parens{1} $
  to the right-hand side. If $a=0$, these terms eventually cancel out
  under refinement and do not affect the fine-partition limit. If $a \neq 0$, 
redefining $\Delta$ by the addition of $a \scalar$ would
  cancel out  the first term. 
  In units where $\hbar$ is not $1,$ this addition 
 is actually $a\scalar \hbar^2$ and thus is a
  quantum correction to the Hamiltonian. This correction presumably
  would correspond
  to a different resolution of the operator-ordering ambiguity
  inherent in promoting
  $g_{ij}p^ip^j$ to an operator.
\end{remark}

\subsection{Convergence and the limit of approximate path integrals} \label{ss:convergence}
Having confirmed that Feynman's time-slicing prescription, adjusted to
satisfy the heat equation for the correct Laplacian up to errors of order
$t^{1/2}$ in the $t$-norm, leads to an
approximate kernel $K_\Delta$, it remains to check that the
corresponding approximate path integrals converge to a definite
limiting kernel, and to check this limit is the heat kernel. Let
$P=(t_1, t_2,\ldots, t_k)$ be a partition, of size $\abs{P}$, of
$t > 0$; that is, $t_i>0$, $\sum_i t_i=t$ and $\abs{P}=\max_i t_i.$
Then the time-sliced approximate path integral based on $K_\Delta$ at
partition $P$ is, suppressing the spatial variables, 
\be
K_\Delta^{*P}(t) = K_\Delta(t_1) *  K_\Delta(t_2) *  \cdots  *  K_\Delta(t_k) .
\ee{k*p-def}
This approximation should get better as the partition $P$ becomes
finer in the following precise sense:  If $P$ is a  partition of $t$ and
 $P'$ is a partition  of $t',$ then the concatenation $PP'$ is a
partition of $t+t';$  if $P_i$ is a partition of $t_i$ for
$1 \leq i \leq k,$ then the partition $P_1P_2 \cdots P_k$ is a
 \emph{refinement} of
 $P=(t_1, \ldots, t_k).$  Note that, by definition, if $Q$ is a refinement of
 $P$ then $\abs{Q} < \abs{P}$. 
\subsubsection{Convergence of approximations to the path integral
based on an approximate semigroup} 
The convergence of approximate path integrals will follow from a Cauchy
property, which says that once $P$ is fine enough, all approximate
path integrals defined by finer partitions $Q$ stay close to the one
defined by $P$. Precisely, suppose $K(t)$ is an approximate semigroup
with constants $B, C, D$ and $T$. The required Cauchy property would
say there is an $A>0$ depending on  
$B, C$ and $D$ such that, if $T$ is chosen small enough,
   \be 
   \norm{K^{* Q}(t)-K^{* P}(t)}_{\tnm{t}}< At^{5/4} \abs{P}^{1/4} 
   \ee{refinement}
for all  refinements $Q$ of all partitions $P$ of $t<T.$
To derive this Cauchy property, first notice Eq.~\ref{eq:k*k-est}, the
defining property of an approximate semigroup, 
readily extends to a version with three terms
\[
\norm{K(t_1) *  K(t_2) *  K(t_3)-K(t)}_{\tnm{t}}\leq
c t^{3/2}.
\]
Write $t = t_1 + t_2 + t_3$, with $0 \leq t_i \leq t/2$ for
$i = 1, 3$ and choose partitions $Q_i$ of $t_i$ again for $i= 1,3$ so that
$Q= Q_1 (t_2) Q_3$. Then
\begin{align*}
  & \norm{K^{* Q}(t)-K(t)}_{\tnm{t}} = \\
  & \qquad \l|\!\l|\sbrace{K^{* Q_1}(t_1)-K(t_1)} *  K(t_2) *
       K(t_3) + K(t_1) *  K(t_2) *  \sbrace{K^{*
           Q_3}(t_3)-K(t_3)}\right. \right. \\
     &\qquad \left.\left. +\sbrace{K^{* Q_1}(t_1)-K(t_1)} *  K(t_2) *  \sbrace{K^{*
           Q_3}(t_3)-K(t_3)} + K(t_1) *  K(t_2) *
       K(t_3)-K(t)\r|\!\r|_{\tnm{t}}.\\
\end{align*}
Induction on the number of entries in $Q$, combined with
Eqs.~\eqref{eq:t*-bd} and~\eqref{eq:k-bd} and a choice of $T$ small
enough that for given $b_1$ and $c_1$ (coming from these inequalities)
$c_1e^{b_1 t^{1/2}}t^{3/2}$ is less than $(1-2^{-1/4})$ leads to
\be
\norm{K^{* Q}(t)-K(t)}_{\tnm{t}} \leq c_1 e^{b_1 t^{1/2}}t^{3/2} ,
\ee{kqk-est}
for some positive constants $c_1$ and $b_1$. Now write
\[
\norm{K^{* Q}(t)-K^{* P}(t)}_{\tnm{t}} =
     \norm{K^{* Q_1}(t_1) *  K^{* Q_2}(t_2) *  K^{* Q_3}(t_3)-K^{*
         P_1}(t_1) *  K(t_2) *  K^{* P_3}(t_3)}_{\tnm{t}},
     \]
for $P = P_1(t_2) P_3$ analogous to the refinement of $Q$ above and
 $Q = Q_1 Q_2 Q_3$ where $Q_2$ is any partition of $t_2$. Another
induction argument, similar to and using the previous result, verifies
Eq.~\eqref{eq:refinement}.

Eq.~\eqref{eq:refinement} says that
for any sequence of partitions $P_1 = (t), P_2, \ldots$ for
sufficiently small $t$, with each partition a refinement of the
previous and with $\abs{P_i} \to 0$ as $i \to  \infty$, $K^{*P_i}(x,y,t)$  is a Cauchy
sequence in the $t$-norm. Eq.~\eqref{eq:tinf-bd} relating the norms
guarantees this sequence is Cauchy in the supremum norm and so by
completeness converges to some $K^\infty(x,y;t)$. Then, for any
partition $P$ of $t$, 
\[
\norm{K^{*P}(t)-K^\infty(t)}_{\tnm{t}} =
           \norm{K^{*P}(t)-K^{*P'_i}(t)  +
          K^{*P'_i}(t)-K^{*P_i}(t)  +
           K^{*P_i}(t)-K^\infty(t)}_{\tnm{t}},
\]
for any sequence $P'_i$ of common refinements of $P$ and
$P_i$. The triangle inequality for the $t$-norm, two applications of
Eq.~\eqref{eq:refinement}, and the convergence of the $K^{*P_i}(t)$ give
\be
  \norm{K^\infty(t) - K^{* P}(t)}_{\tnm{t}} \leq  A t^{5/4}\abs{P}^{1/4}
\ee{kinf-test}
Using the relation between norms, and some judicious choices of
partitions related to $P$ through refinement, this leads to
\be
 \norm{K^\infty(t) - K^{* P}(t)}_{\infty} \leq
 A_1te^{B_1t} \abs{P}^{D_1}
\ee{kinf-est}
for some set of constants $A_1, B_1, D_1$ and $T_1$ (depending on the
previous constants and the dimension $m$) and for all $P$ with $\abs{P} <
T_1$ and all $t$. The $t \abs{P}^{D_1}$ dependence arises from expressing the
change from $P$ to a refinement $Q$ (Eq.~\eqref{eq:refinement}) as a sequence of smaller changes
chosen to exchange the $t$-dependence in the bound in
Eq.~\eqref{eq:tinf-bd}  (describing the relation between the $t$ and
supremum norms) for a
combined $t$ and $\abs{P}$-dependence. The $e^{B_1t}$ ultimately
derives from the bound on the operator norm in terms of the $t$-norm appearing in
Eq.~\eqref{eq:top-bd}, and a bound, for  appropriate partitions $Q$, of
the form
\be
\norm{K^{*Q}(t)}_{(t)} 
\leq e^{bt^{1/2}}
\ee{kq-t} coming from $K^{*Q} = K^{*Q} - K + K$ and Eq.~\eqref{eq:kqk-est}.

Taking $\abs{P}$ to $0$ shows
  \be
  K^\infty(t) = \lim_{\abs{P} \to 0} K^{*P}(t)
  \ee{kinf-lim}
That is, the limit under successive refinements of the approximate
  path integrals based on the approximate semigroup $K$ does
  exist. Therefore $K^\infty(t)$ provides a rigorous definition for
  the path integral based on a first approximation $K$ which may be chosen, as
  above, to be compatible with Feynman's time-slicing prescription for
  a given generalized Laplacian. Note the argument depends only on
  choosing $K$ to be an approximate semigroup; however, for generic
  choices of $K$, the limit $K^\infty$ would not have an interpretation as a path
  integral, as the successive approximations would not correspond to
  time-slicing in any sense. 
\subsubsection{When the approximate semigroup is an approximate heat
kernel, the limiting kernel is the heat kernel}
If $K$ is indeed an approximate heat kernel, such as the specific choice
$K_\Delta$ above coming from time-slicing, it
is necessarily an approximate semigroup, so the approximations $K^{*P}$ to the
path integral  will converge to a path integral
$K^\infty$. The question remains how  $K^\infty$ relates to the heat
kernel for the generalized Laplacian associated to $K$.

Most of the answer follows from allowing the kernel $K^\infty$, for $K$
any approximate heat kernel on a tame vector bundle $\XX$, to act as a
distribution on a sections $f$ of $\XX$. This is
$f(t) = K^\infty(t) * f$. Consider first the small-$t$ limit of
$f(t)$. If $K^\infty(t)$ is 
to agree with the heat kernel, as a distribution, then this limit
should just be $f$. Assuming $f$ is smooth and
bounded on each coordinate patch,  using the trivial
partition in Eq.~\eqref{eq:kinf-test} and then Eq.~\eqref{eq:top-bd} to
relate  the $t$-norm to the operator norm gives,  for $t < T$,
\[
\norm{f(t)- K(t) *  f}_\infty \leq 
   At^{3/2}e^{2B\sqrt{t}}\norm{f}_\infty
\]
from which it follows that $f(t)$ satisfies the correct initial
condition.

To see why the heat equation holds, first note the terminology
``approximate semigroup'' is accurate in that the limiting kernel $K^\infty$, or
in other words the path integral, is a semigroup:  $K^\infty(t)=K^\infty(t_1) *  K^\infty(t_2),$  for $t = t_1 + t_2$
 and $t_1, t_2 > 0$. This follows immediately from considering the limit of
 $K(t_1) * K(t_2)$ under refinements of the partition $(t_1,t_2)$ of
 $t.$ That means in particular $f(t + \tau) = K^\infty(\tau) * f(t)$
 for small $\tau$. Using this in the definition of the $t$-derivative,
 \begin{align*}
     \abs{\frac{\partial f(t)}{\partial t} - \frac{1}{2}\Delta f(t)} &= \abs{\lim_{\tau \to 0}
     \frac{K^\infty(\tau) *  f(t)-f(t)}{\tau} - \frac{1}{2}\Delta f(t)}\\
     &\leq \abs{\lim_{\tau \to 0}
     \frac{K(\tau) *  f(t) -f(t)}{\tau} -
     \frac{1}{2}\Delta f(t)} +
       \lim_{\tau \to
       0} \frac{A\tau^{3/2}e^{2B\sqrt{\tau}}\norm{f(t)}_\infty}{\tau}.
 \end{align*}
That the first term is $0$ is the requirement of Eq.~\eqref{eq:kf} in
the definition of an approximate heat kernel, so $f(t)$ is the
unique solution of heat equation with the initial condition $f$. Thus,
the path integral $K^\infty(t)$ agrees with the heat kernel as a
distribution. Technically, this is only true for $t < T$. However,
larger $t$ can be partitioned as some $Q = (t_1, \dots, t_k)$ with each
$t_i < T$, and then the semigroup
property ensures $K^\infty(t) = (K^\infty)^{*Q}(t)$. Thus, $K^\infty$ is a distributional
   heat kernel for all $t>0.$
Finally, since $\Delta$ is
 elliptic,  standard results on elliptic regularity \cite{Evans98}
 imply $K^\infty(x,y;t)$  is smooth in $x,$ $y,$ and $t$ and
 thus is the heat kernel of $\Delta.$

 That is, the limit of the kernel
 products of any approximate heat 
 kernel is well-defined and agrees pointwise with the
 heat kernel.
 In particular, the choice $K_\Delta$ of
 Eq.~\eqref{eq:kgen-def} shows Feynman's time-slicing prescription
 applied to the action of Eq.~\eqref{eq:action-general} leads to a
 well-defined path integral over paths with fixed endpoints, and this
 path integral is equal to the heat 
 kernel. 
\section{The Atiyah-Singer index theorem for the twisted Dirac
  operator from the twisted $N=1/2$ SUSYQM  path integral}
\subsection{The heuristic argument}
Let $H_{\Delta^{\hat{\VV}}}(x,y, \psi_x, \psi_y, \eta_x,
\eta_y ; t)$ denote the heat kernel for the twisted Dirac operator of
Eq.~\eqref{eq:dirac-def} on sections of the bundle $\hat{\VV}=\spinor
\tensor \Lambda \tspinor$.  The relevant supertrace of the heat kernel is
\[
\str H_{\Delta^{\hat{\VV}}} = \int H_{\Delta^{\hat{\VV}}}(x,x, \psi, \psi, \eta,
\eta ; t) \, d\psi dx.
\]
Now-standard arguments due to
Witten\cite{Witten82a}
in the language of supersymmetry and McKean \&
Singer\cite{MS67} in the mathematics literature say this supertrace,
which is a sum over the eigenvalues but with those on odd-degree
subspaces counting negative, 
computes the index of the Dirac operator.\footnote{This would be the
  Dirac operator on $\hat{\VV}$. For 
  that on $\spinor \times \tspinor$, restrict the heat kernel to the degree-one piece in
  $\tspinor$ and take the trace in $\End(\tspinor)$.} The reason is that the
operator $(\dirac_+ + \dirac_-)$ provides an isomorphism between
even-degree and odd-degree eigenspaces for non-zero eigenvalues. Thus,
in the supertrace these contributions cancel, leaving only $\dim \ker
\Delta_+ - \dim \ker \Delta_-$, which is the index. Notice the index
depends only the heat kernel along the diagonal, and, indeed, only on
its degree-$m$ component as a form on $M$. 

The heuristic Property 1 of the introduction would imply the path integral with the action $S_{\twisted}$ of Eq.~\ref{eq:susyqm-action} and spinor paths
going from $(y, \psi_y, \eta_y)$   to $(x, \psi_x, \eta_x)$  in time
$t$ agrees with
$H_{\Delta^{\hat{\VV}}}(x,y, \psi_x, \psi_y, \eta_x, \eta_y ; t)$.
(The results of Sect.~\ref{ss:convergence} say this is in fact true
for the rigorous path integral $K_{\Delta^{\hat{\VV}}}^\infty$.) With
this, $\str H_{\Delta^{\hat{\VV}}}$ is the path integral taken over
loops.

Now  Property 2 would say the steepest descent approximation gives the small-$t$ behavior of the path integral
over loops. Since the index does not depend on $t$, this small-$t$
approximation computes the index as the integral over $(x,
\psi)$ of the steepest descent approximation to path integral on
loops based at this 
point. The resulting equation expresses the topological index as an
integral over $M$, the integral over $\psi$ serving to
pick out the top-form piece. This is the index theorem. 

To compute the steepest descent approximation, expand the action  $S_{\twisted}$ about
its minimum, after rescaling the paths
according to their expected contribution for small-$t$, and discard
terms of  order higher than $1$ in $t$. The result is an approximate action
$S_{q}$ which is purely quadratic in the paths. The path integral
taken over based loops with 
action $S_q$ then reduces heuristically to ratios of the determinants of the
differential operators appearing in $S_q$. 
\subsection{A rigorous version of rescaling}
\subsubsection{Reduction in a neighborhood of the diagonal to a
  trivial bundle in $\RR^m$} \label{sss:diagonal-neighborhood}
Work locally in the bundle $\VV =\spinor \tensor \Lambda \tspinor$
over $M$ (henceforth dropping the hats). Let $x_0 \in M$.  Endow a ball of radius $D_1>0$ around $x_0$  with Riemann
normal coordinates, and identify the restriction of $\VV$ to this ball
 with $\VV_{x_0}$
 via parallel transport along minimal geodesics.  This defines a
 metric
 $g_1,$ a trivial bundle $\VV_1,$ and a connection $\Nabla^1$ over a
 neighborhood of the origin in $\RR^m,$ all with bounded derivatives
 up to order four.  Extend all of these to all of $\RR^m$ so that the
 derivatives remain bounded and so that both $\Nabla^1$ and the
 Levi-Civita connection $\Nabla^{g_1}$ continue to be $0$ on radial
 directions. Let $\Clifford$ denote the Clifford algebra  $C(T^*_{x_0}M)$ at
 $x_0 = 0,$
 whose action on $\VV_{x_0}$ splits it into $\spinor
 \tensor \tspinor,$ where $\spinor$  is the spinor representation of
 $\Clifford$ and $\Clifford$ acts trivially on $\tspinor$. $\VV_1$ can be identified with the trivial
 bundle $\spinor \tensor \tspinor$ over  $\RR^m$.
  Identifying the
 Clifford algebra   at any point in $\RR^m$  with $\Clifford$ by
 radial translation gives it an action on $\spinor \tensor
 \tspinor$ that
 makes $\Nabla^1$ a Clifford connection  agreeing 
 with $\Nabla^\VV$
 in the ball of radius $D_1$.  In fact then $\Nabla^1 = \Nabla^{g_1}
 \tensor 1 + 1 \tensor \Nabla^{\tspinor},$ where $\Nabla^{g_1}$ is the
 Levi-Civita connection on $\spinor$ and $\Nabla^\tspinor$ is some
 connection on $\tspinor$ with curvature $F^\tspinor$.  The choice $V_1=c(F^\tspinor)-
 \scalar_1/4$ defines
 a Dirac operator $\dirac_1$ on $(g_1,\spinor\tensor
 \tspinor\times \RR^m, \Nabla^1)$, and a generalized Laplacian
 $\Delta_1 = (\dirac_1)^2$, whose associated approximate heat kernel
 $K_1=K_{\Delta_1}$
 can be identified with $K_{\twisted}$ of Eq.~\eqref{eq:ksusy} in that ball
 by the obvious isomorphism. 

 On the other hand, given any pair of approximate semigroups, each on
 a tame bundle over a tame Riemannian manifold, if there is an bundle
 isomorphism respecting the tameness structure under which the two
 semigroups are identified (via pullback) in some neighborhood of a
 given point,  then the corresponding path
 integrals will agree on the diagonal at that point up to terms that are
 exponentially damped as $t \to 0$. That is, letting $\Phi$ denote
 the isomorphism, there are real positive constants $c$, $d$ and $T$,
 depending on the constants $B_i$ $C_i$, and $D_i$ for $i= 1,2$
 required to define the semigroups, such that
 \be
\abs{K_1^\infty(x,x;t) -K_2^\infty(\Phi(x),\Phi(x);t)} \leq ce^{-d/t}.
\ee{local}
The argument for this starts by breaking $P$ up according to its
  intervals as $P = P_j(t_j)P_{j'}$ and writing $K_1^{*P} -
  K_2^{*P} = \sum_j K_1^{*P_j} *  K_1(t_j) * K_2^{*P_{j'}} -
  K_1^{*P_j} *  K_2(t_j) * K_2^{*P_{j'}}$. If the left-hand side is
  being evaluated at a pair of points on which the two semigroups
  agree,  the equation is unaffected by adding the assumption that in
  each term the lone 
  $K_1$, and hence the preceding $K_1^{*P_j}$, is evaluated at $(y_{j-1},y_j;t_j)$ for $y_{j-1}$ outside the neighborhood
  of agreement and $y_j$ inside, while  in $K_2(\Phi(y_{j-1}), \Phi(y_j);
  t_j)$ the point $y_{j-1}$ is inside and $y_j$ outside this
  neighborhood. With this added assumption, a bound analogous to that of
  Eq.~\eqref{eq:kq-t} on the growth of kernel products in the
  $t$-norm for a
  semigroup and Eq.~\eqref{eq:tpntws-bd} relating the $t$-norm to the
  supremum norm lead to $\abs{K_1^{*P}(x,x;t)
    -K_2^{*P}(\Phi(x),\Phi(x);t)} \leq c_1e^{-d_1/t}$. The convergence
  result for semigroups, specifically Eq.~\eqref{eq:kinf-test},
  readily gives the agreement in the path integrals on the diagonal.

  The upshot is that to understand the behavior, on the diagonal for
  short times,  of the path integral
  based on the approximate kernel
  $K_{\twisted}$ it suffices to understand that of the path 
  integral based on $K_1$ in the simpler setting of the trivial
  bundle $\VV_1$ on $\RR^m$.
\subsubsection{Rescaling the local kernel} \label{sss:rescale}
 Because $\VV_1$
 is trivial, $K_1$ can be taken to be a function on $\RR^m \times
 \RR^m$ with
 values in $\End\parens{\spinor} \tensor \End\parens{\tspinor} \sim \Clifford
 \tensor \End(\tspinor)$.
The Clifford algebra 
action $c_{\Lambda}$ on $\Lambda T^*_{x_0}M$ maps $K_1$ to a function
with values in $\End\parens{\Lambda
  T^*_{x_0}M}\tensor \End(\tspinor)$. Mildly abuse notation to let $K_1$
also refer to this function. 

Rescale by a parameter $0 \leq r \leq 1$ as follows: Define $\phi_r \colon \RR^m
 \to \RR^m$ by $\phi_r(\vec{x})=r\vec{x}$,  and define $\psi_r \colon
 \Lambda T^*_{x_0}M \to \Lambda T^*_{x_0}M$ for elements $\alpha$ of a
 given degree by $\psi_r(\alpha)=
 r^{\deg(\alpha)} \alpha$. For the metric, define
 $g_r = r^{-2} \phi_r^*[g_1]$. This extends continuously to
 $g_0=g_{1,\vec{0}}$, where, by construction, $g_{1,\vec{0}} (\vec{v},
 \vec{w}) = \parens{\vec{v}, \vec{w}}$, 
 the standard inner product on $R^m$.
 Finally, for  $K(\vec{x},\vec{y};t)$  a kernel on the bundle $\Lambda T^*_{x_0}M
 \times \tspinor$ over $\RR^m,$ define
 \be \Phi_r[K](\vec{x},\vec{y};t) = r^{m} \psi_r^{-1}
 K(r\vec{x},r\vec{y};r^2t) \psi_r.
\ee{rescale}
Write 
 \[K_r=\Phi_r(K_1) \]
 for the rescaled version of $K_1$.

 The family of metrics has the properties  (extending each
 formula by continuity to $r=
 0$):
 \begin{align*}
	g_{r,\vec{x}}(\vec{v}, \vec{w}) & = g_{1,r\vec{x}}(\vec{v}, \vec{w})  \\
	d_{g_r}(\vec{x},\vec{y}) & =  r^{-1}d_{g_1}(r\vec{x}, r\vec{y} )    \\
	\parens{\vec{y}_\vec{x}}_{g_r}  & =  r^{-1} \parens{(r\vec{y})_{r\vec{x}}}_{g_1} \\
	\Ricci_r(\vec{y}_{\vec{x}},\vec{y}_{\vec{x}}) & =
        \Ricci_1\parens{(r\vec{y})_{r\vec{x}},(r\vec{y})_{r\vec{x}}}
        \\ 
	\scalar_r & = r^2 \scalar_1\\
	\dmeas_{g_r}\vec{y} & = r^{-m} \dmeas_{g_1} (r\vec{y}).
 \end{align*}

 Direct calculation shows the rescaling commutes with the kernel
 product. Further, as the constants $B_1$  and $D_1$ in the definition
 of an approximate semigroup and the $t$-norm depend only on the
 supremum of the metric $g_1$, and the rescaling from $g_1$ to $g_r$
 cannot change the supremum, these constants will work for any of the
 $g_r$, in the sense that there is a $t$-norm which is independent of $r$. In
 fact, by making $D$ a fixed fraction of $D_1$, a straightforward
 argument shows there are constants such that for all $0 < r \leq 1$,
 $\norm{\Phi_r(K)}_{\tnm{t}} \leq \norm{K}_{\tnm{t}}$. From this,
 directly checking the definition shows $K_r$ is an approximate
 semigroup, so the path integral $K^\infty_r$ based on $K_r$ will be
 well-defined. Further, since rescaling commutes with the kernel products
 defining the approximate path integrals,
 \be
 K_r^\infty = \Phi_r\parens{K_1^\infty}.
 \ee{scaling-interchange}
That is, rescaling the path integral based on the approximate kernel $K_1$, which
up to exponentially-damped terms agrees on the diagonal with the heat
kernel for $\Delta^\VV$, gives the same result as basing the path
integral on the approximate kernel $K_r$.

If this extends to $r=0$, it will say any aspect of the heat kernel
(on the diagonal) which can be calculated from the $r=0$ limit of the
rescaling applied to the path integral can in fact be calculated by a presumably
simpler path integral based on the $r=0$ limit of $K_r$. There will still be
two issues:
\begin{enumerate} \item Does the rescaling limit of the path integral retain enough
  information to compute the supertrace?
  \item Does the rescaling limit of
$K_r$ lead, via the refinement limit of its products,  to a computable path integral?
\end{enumerate}
Address the second question first, by considering what happens to
 \begin{align*}
K_r = &\lim_{r \to 0} r^m (2\pi t)^{-m/2} e^{-\sbrace{d_{g_r}(\vec{x},
      \vec{y})}^2/(2t)} \\ 
&\qquad \times e^{-\Ricci_r(\vec{y}_\vec{x},\vec{y}_\vec{x})/12  + t
          \scalar_r / 24 - \frac{t}{4} F_{ij}^\tspinor(r\vec{x})\psi_r^{-1}
          r^2 c(dx_i)c(dx_j)\psi_r} \psi_r^{-1}
        \pt_{r\vec{x}}^{r\vec{y}} \psi_r 
\end{align*}
as $r \to 0$. Since $g_1= g_0 + \OO\parens{\abs{\vec{x}}^2}$ both
curvature terms vanish in the limit, and
$d_{g_r}(\vec{x}, \vec{y}) \to \abs{\vec{x}-\vec{y}}$.
Direct calculation shows $\lim_{r\to 0} \psi_r^{-1}rc(dx)\psi_r=dx$, so 
\[
\lim_{r\to 0} F_{ij}^\tspinor(r\vec{x})\psi_r^{-1}
r^2 c(dx_i)c(dx_j)\psi_r/2 = \Flimit,
\]
where
\[
\Flimit=\frac{1}{2}F_{ij}^\tspinor(x_0)
dx^i \wedge dx^j
\]
defines $\Flimit$ as an element of $\Lambda T^*_{x_0}M
\tensor  \End\parens{\tspinor}$ (that is, a  $2$-form at $x_0$
taking values in linear transformations on the vector space
$\tspinor$). Finally, in $\psi_r^{-1} \pt_{r\vec{x}}^{r\vec{y}}
\psi_r$,  with the bundle being
  trivialized radially at the origin, the parallel
  transport from $r\vec{x}$ to $r\vec{y}$ is the holonomy of the
  geodesic triangle from $0$ to $r\vec{x}$ to $r\vec{y}$ to $0$.
    In $\tspinor$, this holonomy differs from $1$
  by a quantity proportional to the area enclosed, which is
  $\OO(r^2)$ and will thus vanish in the limit. 
  For the  $\Lambda(T^*_{x_0}M)$ piece, the holonomy is an element
  of the spin group and therefore an exponential of a degree-two
  element of $\Clifford$. 
  This Clifford element
  in turn is the image under $c$ of
  the two-form generating
  the holonomy about the same geodesic triangle with respect to the
  Levi-Civita connection. Standards results \cite{AS53} say this
 is $\parens{\Rlimit \cdot r\vec{x}, r\vec{y}-r\vec{x}}/4
  + \OO\parens{\abs{r\vec{x}} \abs{r\vec{y}-r\vec{x}}
    \abs{r\vec{y}+r\vec{x}}}$, where analogously to $\Flimit$,
  \[
  \Rlimit_k^l= \frac{1}{2}R_{ijk}^{l}(x_0)  dx^i \wedge dx^j
  \] 
defines
 $\Rlimit \in \Lambda T^*_{x_0}M \tensor \End(T_{x_0}M)$.
  Thus, this piece 
  is the exponential of
  the image under $c$ of $\parens{\Rlimit\cdot  r\vec{x},r\vec{y}}/4+
  \OO\parens{r^3}$. Conjugation by $\psi_r$ will
  reduce the power of $r$ by two, giving
\[
   \lim_{r\to 0} \psi_r^{-1}
        \pt_{r\vec{x}}^{r\vec{y}} \psi_r=e^{\parens{\Rlimit \vec{x},
            \vec{y}-\vec{x}}/4}.
\]
Putting this all together,
\[
\lim_{r \to 0} K_r = K_0,
\]
where
\be
 K_0(\vec{x},\vec{y};t) = (2\pi t)^{-m/2}
e^{-\abs{\vec{y}-\vec{x}}^2/(2t)}
 e^{\parens{\Rlimit \vec{x}, \vec{y}-\vec{x}}/4 -t\Flimit/2}.
\ee{k0-def}
The kernel $K_0$ is in fact a time-slicing approximation
for the path integral with action $S_q$, as anticipated by the
heuristic application of steepest descent.  Using Lebesgue Dominated
Convergence it is straight-forward to show that for an fixed partition $P$ of any $t>0$
\[
\lim_{r \to 0} K_r^{*P}(0,0;t)=K^{*P}_0(0,0;t).
\]
That is, approximate path integrals based on $K_r$ go to those based
on $K_0$ in the rescaling limit. The starting point is to observe
$K_0$ and $K_r$ are bounded by
$C_1 H(\vec{x},\vec{y};C_2t)$ for some $C_1,C_2$ where
$H(\vec{x},\vec{y};t) = (2\pi t)^{-m/2}
e^{-d_{g_0}^2(\vec{x},\vec{y})/(2t)}$,
which in turn follows from the same bound on $K_1$. 

Unfortunately, the appearance of $R\cdot \vec{x}$ in the exponential,
and the fact that $\vec{x}$ is free to range over all of $\RR^m$ 
though $\abs{\vec{x} - \vec{y}} < D$, means that $K_0(t)$ will not satisfy the
definition of an approximate semigroup, so the preceding convergence
results do not immediately apply to ensure the refinement limit
$K_0^\infty$, and
hence the path integral even exist.
On the other hand, it should be a standard
result on path integrals with quadratic actions that the path integral
based on $K_0$ is well-defined and agrees with the heat
kernel for a Laplacian compatible with the Lagrangian whose action is
$S_q$. For a proof in the language of kernel products
 see
\cite{FS17}. The precise statement is
\be
	\lim_{\abs{P} \to 0} K_0^{*P}=K^\infty_0
\ee{k0-inf}
	converges pointwise, and is the heat kernel for the operator
\be
	\Delta= \frac{\partial^2}{\partial x_i \partial x_i} +
        \frac{1}{2}\Rlimit_i^j x_j \frac{\partial}{\partial x_i
        }- \Flimit + \abs{\Rlimit \cdot \vec{x}}^2/16.
\ee{k0-delta}
Standard results on heat kernels for this Laplacian, which 
physically is just the Hamiltonian for a particle in a constant magnetic
field, give the explicit value on the diagonal:
	\be
	K_0^\infty(0,0;t)= (2\pi t)^{-m/2}
{\det}^{1/2}\parens{\frac{t\Rlimit/4}{\sinh(t\Rlimit/4)}} e^{-t\Flimit/2}.
\ee{k0-form}
The ratio of determinants on the right-hand side is that predicted by
the  heuristic path integral with action $S_q$. 
  \subsubsection{Along the diagonal, the small-$t$ behavior of the
    full path integral agrees with that of the rescaling limit}
Return now to the question of whether the rescaling limit captures
enough of the small-$t$ behavior to compute the supertrace of the full
heat kernel (Question 1 above). Working directly from the heat
equation, it is easy to 
see that $K^\infty_{1}(0,0;t)= \sum_{i=0}^{(m+2)/2} A_i t^{i-m/2} +
\OO\parens{t}$ for $A_i \in \Clifford \tensor \End(\tspinor)$, where
each $A_i$ is of degree $2i$ in the Clifford filtration. Thus,
$c_\Lambda(A_i)$ is of degree at most $2i$ as an element of
$\End(\Lambda T^*_{x_0}M) \tensor \End(\tspinor)$. The supertrace,
which includes an integration over $M$, will pick out the 
degree-$m$ piece of $K^\infty_{1}(0,0;t)$. This piece comes from $A_{m/2} +
A_{(m+2)/2} t + \OO\parens{t}$. As $t$ goes to $0$, the supertrace 
thus sees only $c_\lambda \parens{A_{m/2}}$, and of that, only the 
piece of degree $m$.  

To see this is also exactly the piece that survives the rescaling limit, first apply the rescaling, which takes the term $c_\lambda(A_i)
t^{i-m/2}$ to $\psi_r^{-1} c_\Lambda(A_i) \psi_r
r^{2i}t^{i-m/2}$. Since $c_\lambda(A_i)$ is a sum of terms of  degrees
up to $2i$, and conjugation by $\psi_r$ multiplies a term of degree
$k$ by $r^{-k}$, the result is an overall factor of $r^{2i-k}$. As $r
\to 0$, only the ``top'' piece, of degree $2i$, will survive. Moreover, the last
term in the series, where $i= m/2 + 1$ and so the top piece of $A_i$
has degree $m + 2$, will go to $0$ after rescaling and taking $r \to
0$. Likewise, the $\OO\parens{t}$ corrections vanish in this rescaling
limit. In short, 
\[
\lim_{r \to 0} \Phi_r\sbrace{K_1^\infty}(0,0;t)= 
\sum_{i=0}^{m/2}\rho\parens{A_i}t^{i-m/2},
\]  
where $\rho$ takes an element $A_i$ of Clifford degree $2i$ to the
form of degree $2i$ corresponding to the top-form piece of
$c_\lambda(A_i)$. In particular, the degree-$m$ piece of the rescaling
limit of the path integral agrees with that of the small-$t$ limit of
the heat kernel (path integral)
$K_1^\infty(0,0;t)$, so the rescaling limit indeed captures enough of
  the full heat kernel to calculate the small-$t$ limit of the
  supertrace.
\subsection{The index theorem}
  It remains to check  the rescaling limit of the path
  integral  is the same as the path integral based on the rescaling
  limit of the approximate heat kernel; that is, to check
  \[
\lim_{r \to 0} \Phi_r[K_1^\infty](0,0;t)=  K_0^{\infty}(0,0;t).
  \]
For fixed $t$,  there is a choice of $P$ making both
$K_0^{*P}(0, 0; t)$ arbitrarily close to $K_0^{\infty}(0,0;t)$ and
$K_r^{*P}(0, 0; t)$ arbitrarily close to $\Phi_r[K_1^\infty](0,0;t)$,
for {\em all} $r \in (0, 1]$\footnote{This relies on the fact that the
  constants making $K_r$ an approximate semigroup do not depend on
  $r$.}. With $P$ fixed, there is a choice of $r$ making $K_r^{*P}(0,
  0; t)$ close to $K_0^{*P}(0, 0; t)$. That means
  choosing this $P$ and $r$ combination will make
  $\Phi_r[K_1^\infty](0,0;t)$ arbitrarily close to
  $K_0^{\infty}(0,0;t)$, which is the statement of convergence.

  Putting this all together,
  \[
  \str K^\infty_\Delta(t) = \lim_{t \to 0} \str K^\infty_\Delta(t) =
  \int_M (2\pi t)^{-m/2} {\det}^{1/2}\parens{\frac{t\Rlimit/4}{\sinh(t\Rlimit/4)}}
e^{-t\Flimit/2}.
\]
This is the Atiyah-Singer index theorem for the twisted Dirac
operator. In fact, the  argument says something about the
lower-degree terms in the expansion for the heat kernel; namely, writing
\[P(t) = \sum_{k = 0}^{\infty} A_k t^{k
  -m/2}\]
for the Laurent series in $t$ asymptotic to the diagonal of the heat kernel $K^\infty_{\Delta}(x_0,x_0;t)$ of
$\Delta = \dirac^2,$
\[
\rho\parens{P(t)}=  \sum_{k = 0}^{m/2} \rho(A_k) t^{k
  -m/2} =
(2\pi t)^{-m/2} 
{\det}^{1/2}\parens{\frac{t\Rlimit/4}{\sinh(t\Rlimit/4)}}
e^{-t\Flimit/2} .
\]
This statement is the local form of the index theorem.
\section{Some conclusions}
This completes the work of filling in the details to apply Feynman's
time-slicing prescription to define the path integral for twisted
$N=1/2$ SUSYQM on a Riemannian manifold and to check it has the
properties Witten, Alvarez-Gaum\'{e}, Friedan and Windey assume in
their path integral proofs of index theorems. The definition, and the
representation of the heat kernel as a path integral,  extends
to generalized Laplacians on any vector bundle. The  method
of proof appears to single out a
particular resolution of the operator-ordering ambiguity inherent in
passing from the Lagrangian to a Hamiltonian as one providing faster
convergence of the approximations to the path integral.

This approach to proving the convergence of time-slicing approximate
path integrals may apply to other settings, particularly those where
exact evaluations of the path integral are feasible. These include
two-dimensional Yang-Mills, which in fact is known to reduce to
quantum mechanics~\cite{Fine91}, Chern-Simons theory~\cite{Witten89a}, and
cohomological field theories~\cite{BT93, SW94}. The path integral
arguments in the latter are closely analogous to those for SUSYQM. 

\begin{acknowledgements} 
It is a pleasure to thank the Mathematics Department at MIT for
hosting me as a Visiting Professor while I completed this
work.
\end{acknowledgements}

\def\cprime{$'$}

\end{document}